\documentclass[aps, preprint, a4paper, nofootinbib, longbibliography]{revtex4-1}

\usepackage{epsfig}
\usepackage[top=75pt,bottom=75pt,left=50pt,right=50pt]{geometry}
\usepackage{enumitem}
\usepackage{caption}
\usepackage{amsmath}
\usepackage{amssymb}
\usepackage{upgreek} 
\usepackage{graphicx}
\usepackage{booktabs}
\usepackage{epstopdf}
\usepackage{gensymb}
\usepackage[colorlinks=true,citecolor=blue,linkcolor=blue]{hyperref}
\usepackage[capitalise,noabbrev]{cleveref}
\usepackage{bm} 
\usepackage{tikz,xcolor}

\usepackage[utf8]{inputenc}

\definecolor{lime}{HTML}{A6CE39}
\DeclareRobustCommand{\orcidicon}{\hspace{-4pt}
	\begin{tikzpicture}
		\draw[lime, fill=lime] (0,0) 
		circle [radius=0.16] 
		node[white] {\hspace{0.1mm}{\fontfamily{qag}\selectfont \tiny ID}};
		\draw[white, fill=white] (-0.07,0.1) 
		circle [radius=0.01];
	\end{tikzpicture}
	\hspace{-3.2mm}
}

\foreach \x in {A, ..., Z}{\expandafter\xdef\csname 
	orcid\x\endcsname{\noexpand\href{https://orcid.org/\csname orcidauthor\x\endcsname}
		{\noexpand\orcidicon}}
}

\begin{document}

\title{Probing the cosmic sterile-neutrino background with IceCube}
 
\author{Bhavesh Chauhan\orcidA{}}
\email{bhavesh.chauhan@pilani.bits-pilani.ac.in}
\affiliation{Department of Physics, Birla Institute of Technology and Science, Pilani 333031, India}

\author{Priyank Parashari\orcidB{}}
\email{ppriyank@usc.edu}
\affiliation{Centre for High Energy Physics, Indian Institute of Science, Bengaluru 560012, India}
\affiliation{Department of Physics \& Astronomy, University of Southern California, Los Angeles, CA, 90007, USA}

\date{\today}

\begin{abstract}
In this paper, we take a close look at the interaction between the TeV--PeV energy astrophysical neutrinos and a hypothetical cosmic sterile-neutrino background. These interactions yield absorption features, also called ``dips", in the astrophysical neutrino spectrum, which are studied using the deposited energy distribution of high-energy starting events (HESE) in the IceCube detector. We improve upon the previous analysis by including the effects of regeneration and a realistic source distribution on the propagation of astrophysical neutrinos. We use the latest 7.5-year HESE dataset and include the observation of Glashow resonance in our analysis. We evaluate the impact of these dips on the inferred spectral index and overall normalization of the astrophysical neutrinos. We find a mild preference for dips in the 300--800 TeV range, and the best-fit parameters for the mass of sterile-neutrino and the mediator are 0.5 eV and 23 MeV, respectively. We find that the inclusion of these absorption features lowers the spectral index of astrophysical neutrinos to $2.60^{+0.19}_{-0.16}$. We show qualitatively that the lower spectral index from HESE sample can reduce the disagreement with the Northern Tracks sample. We also forecast the event spectrum for IceCube-Gen2 for the two different fits.
\end{abstract}

\maketitle


\section{Introduction}

The IceCube neutrino observatory has detected neutrinos of astrophysical origin in the TeV--PeV range in several detection channels. The latest 7.5 years (2635 days) high energy starting events (HESE) sample consists of 102 events, of which 60 events have reconstructed deposited energy above 60 TeV \cite{IceCube:2020wum}. The IceCube experiment has also observed a Glashow Resonance (GR) event  \cite{Glashow:1960zz} in the 4.6 years PeV energy partially contained events (\text{PEPE}) sample \cite{IceCube:2021rpz}. IceCube also maintains a catalog of muon-like tracks originating from the northern hemisphere (henceforth called Northern Tracks). A large fraction of the Northern Tracks are below 100 TeV and originate via atmospheric neutrinos. The highest energy events from this sample have astrophysical origin.

It is abundantly clear that these high-energy neutrinos have an extra-galactic origin and are likely to originate in sources capable of accelerating cosmic rays to EeV-scale energies.  The neutrinos are produced in subsequent interactions of these ultra-high energy cosmic rays with ambient gas or radiation. At the time of writing, only a handful of such sources have been confirmed~\cite{IceCube:2018cha,IceCube:2018dnn,Aartsen:2019fau,IceCube:2022der}. If the neutrinos are produced via interactions with ambient gas, i.e. in proton-proton ($pp$) interactions, the neutrinos are expected to have a power-law spectrum \cite{Margolis:1977wt,Stecker:1978ah,Kelner:2006tc}. In the baseline scenario, the neutrinos oscillate and reach flavor equilibrium before they arrive to the Earth \cite{Lipari:2007su,Choubey:2009jq,Mena:2014sja,Mena:2014sja,IceCube:2015rro}. One usually assumes that an isotropic flux of astrophysical neutrinos is incident on the Earth with a total differential all-flavor neutrino-plus-anti-neutrino spectrum given by
\begin{equation}\label{eq:spl}
	\dfrac{d\Phi_{6\nu}}{dE_\nu} = \Phi_{\texttt{astro}} \, \left( \frac{E_\nu}{E_0}\right)^{-\gamma_{\texttt{astro}}}\,\cdot 10^{-18}\, \text{GeV}^{-1}\, \text{cm}^{-2}\,\text{s}^{-1}\,\text{sr}^{-1},
\end{equation}
which is characterized by two free parameters: a spectral index ($\gamma_{\texttt{astro}}$) and the normalization scale ($\Phi_{\texttt{astro}}$). The reference energy scale $E_0$ is chosen to be 100 TeV. If the neutrinos are produced via photo-hadronic interactions, i.e. in proton-gamma ($p\gamma$) sources, the energy spectrum is expected to be ``bump-like" \cite{Kelner:2006tc, Mucke:1999yb,Hummer:2010vx, Fiorillo:2022rft}. Realistically, both $pp$ and $p\gamma$ sources are expected to contribute. In this work, we will assume that the $pp$ sources dominate and the diffuse flux of astrophysical neutrinos is described by a single power-law (SPL), as in \Cref{eq:spl}.       

Are the events observed by IceCube consistent with the assumption of SPL? To a large extent, \emph{yes}. However, there are some tantalizing discrepancies that have motivated a large body of work in the last decade. Some of the explanations are based on dark matter \cite{Feldstein:2013kka, Esmaili:2013gha, Bai:2013nga, Fong:2014bsa, Rott:2014kfa, Esmaili:2014rma, Murase:2015gea, Anchordoqui:2015lqa, Bhattacharya:2016tma, Bhattacharya:2017jaw, Hiroshima:2017hmy, Chakravarty:2017hcy}, resonances beyond standard model \cite{Anchordoqui:2006wc,Barger:2013pla,Dutta:2015dka,Dey:2015eaa,Dev:2016uxj,Mileo:2016zeo,Chauhan:2017ndd,Dey:2017ede,Becirevic:2018uab,Babu:2019vff,Dey:2020fbx}, or non-standard/secret neutrino interactions \cite{Ng:2014pca, Ioka:2014kca, Ibe:2014pja, Blum:2014ewa,Araki:2014ona,Cherry:2014xra,Kamada:2015era,DiFranzo:2015qea,Araki:2015mya,Rasmussen:2017ert,Jeong:2018yts,Mohanty:2018cmq,Pandey:2018wvh,Pandey:2019apj,Bustamante:2020mep,Creque-Sarbinowski:2020qhz,Mazumdar:2020ibx,Fiorillo:2020zzj,Fiorillo:2020jvy,Carpio:2021jhu,Esteban:2021tub}. One such discrepancy is the inferred value of the spectral index in two independent datasets \cite{IceCube:2020wum, Abbasi:2021qfz, Stettner:2019tok, IceCube:2025ary}, 
$$ \gamma_{\texttt{\,astro}}^{\text{\,HESE}} = 2.87^{+0.20}_{-0.19} \quad \text{and} \quad \gamma_{\texttt{\,astro}}^{\text{\,Nor. Tr.}} = 2.37^{+0.09}_{-0.09}\,.$$ 
The HESE sample selects events (cascades, tracks, and double-cascades) that start inside the instrumented detector volume. Any event that deposits its first significant light inside the fiducial volume, without coinciding hits in the outer region, is considered a starting event. Such a strong veto on the outer layers ensures that the event is likely to be induced by a neutrino rather than a through-going muon. Furthermore, an energy cut-off of 60 TeV is considered to reduce the atmospheric neutrinos from the sample. The Northern Tracks sample selects upward-going muon-like tracks that are mostly through-going muons that originate in neutrino interactions outside the instrumented detector volume. This sample has low energy-threshold and very good direction reconstruction, which also allows for point-source searches with this dataset. The Northern Track events above 100\,TeV mostly contain events of astrophysical origin \cite{Abbasi:2021qfz, IceCube:2025ary}. If the two samples originate from the \emph{same} diffuse flux of astrophysical neutrinos, then the $\sim 2.3\sigma$ discrepancy between the two fits to the spectral index is quite puzzling. Interestingly, the Enhanced Starting-Track Event Selection (ESTES) sample, which selects muon-like tracks that start inside the detector irrespective of their direction, reports an intermediate value for the spectral index \cite{IceCube:2024fxo}, 
\begin{equation}\nonumber
    \gamma^{\text{ESTES}}_{\texttt{astro}} = 2.58^{+0.10}_{-0.09}.
\end{equation}
We refer the reader to Fig. 7 in  Ref. \cite{IceCube:2025ary} for a compendium of fits to the parameters of astrophysical neutrino flux from various IceCube datasets.  

The primary aim of this paper is to answer-- \emph{How are these fits affected if there are dips in the astrphysical neutrino spectrum?} Such dips are a consequence of interaction between astrophysical neutrinos and a cosmic background. If the interaction is through an $s$-channel process, then the absorption feature is narrowly located around the resonance energy $E_\nu^{\rm res} = M_X^2/2m_b$, where $M_X$ is the mass of the mediator and $m_b$ is the mass of the background particle. If the interaction is $t$-channel, then a broad spectrum feature is expected \cite{Mohanty:2018cmq}. In Ref. \cite{Esteban:2021tub}, the effect of non-resonant contribution to the process was found to be significant for larger couplings, and we have included them in our analysis. With only standard model interactions, the dip will occur for neutrinos with energy EeV \cite{Weiler:1982qy,Brdar:2022kpu}. However, if there are non-standard/secret interactions mediated by a MeV-scale boson, then the dips can occur in the TeV--PeV range \cite{Ng:2014pca}. In this paper, we study the scenario proposed in Ref. \cite{Chauhan:2018dkd} where the dip is caused by interactions between astrophysical neutrinos and a cosmic sterile-neutrino background.

\begin{figure}[b]
    \centering
    \includegraphics[width=0.5\linewidth]{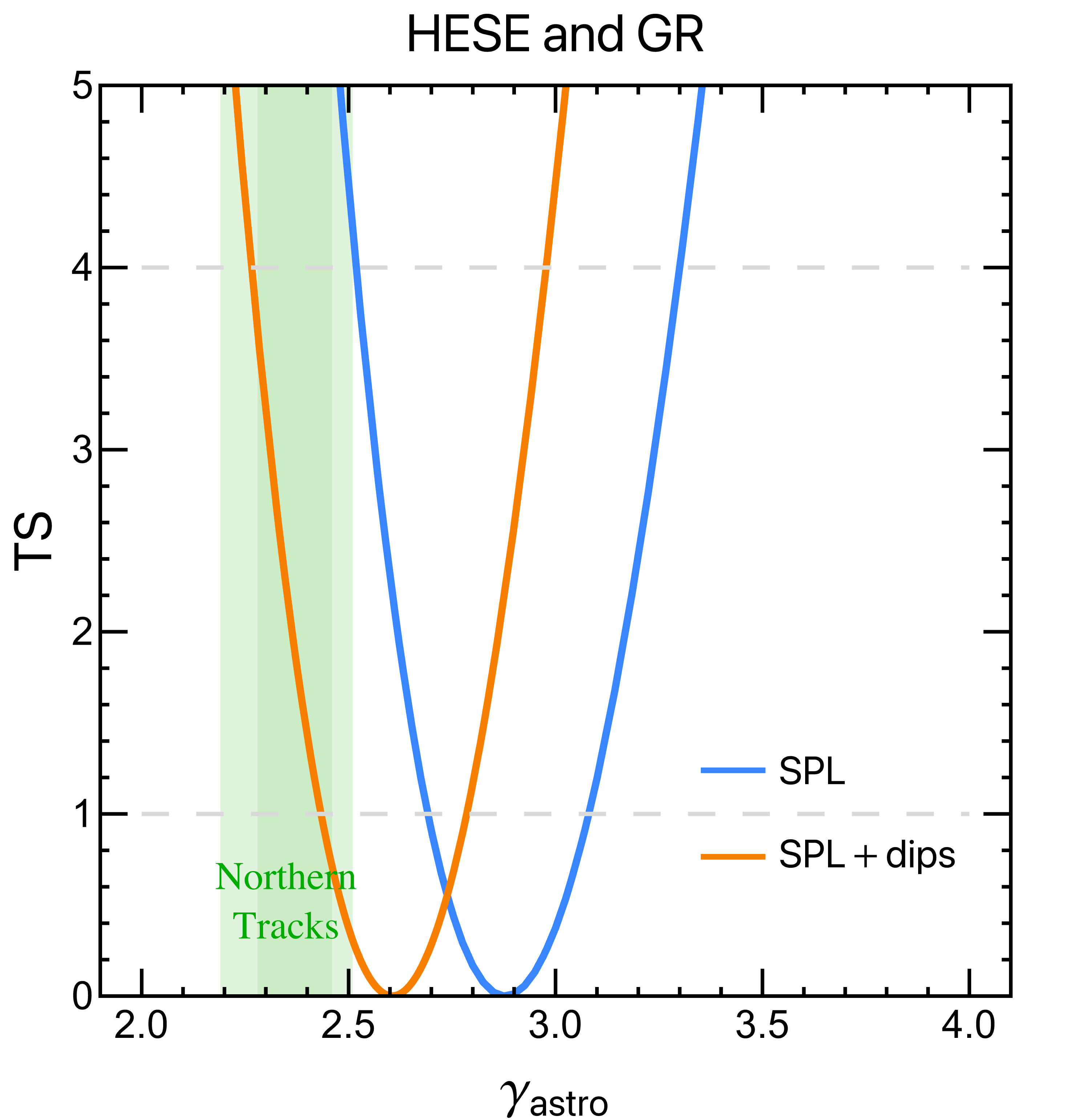}
    \caption{The test statistic (see \Cref{eq:ts}) for the single power-law fit of 7.5 years HESE sample and Glashow resonance \cite{IceCube:2020wum} is shown in blue. In orange, we show the test statistic after including absorption features. The horizontal grey dashed lines are used to denote the 68\% C.L. and 90\% C.L. intervals. We have shown the results from the 11-year Northern Tracks sample \cite{Abbasi:2021qfz} in green.}
    \label{fig:gamma}
\end{figure}

The light sterile neutrino is a well-motivated extension of the standard model and has been the subject of investigation in a number of experiments. Currently, there is evidence in support of ~\cite{Anselmann:1994ar,Hampel:1997fc,AguilarArevalo:2008rc,AguilarArevalo:2010wv,Aguilar:2001ty,Mention:2011rk,Aguilar-Arevalo:2018gpe,Serebrov:2020kmd} as well as against \cite{MicroBooNE:2022sdp} the sterile neutrino. For a recent review, we refer to Ref.\,\cite{Dasgupta:2021ies}. The presence of a fully thermalized sterile neutrino is in conflict with the measurement of effective number of neutrino-like species ($N_{\text{eff}}$) from cosmological probes~\cite{Hamann:2011ge,Cooke:2017cwo,Planck:2018vyg}. If the sterile neutrino has self-interactions via a MeV scale vector mediator, their recoupling can be delayed, and the $N_{\text{eff}}$ constraints can be avoided \cite{Hannestad:2013ana,Dasgupta:2013zpn}. The case for pseudo-scalar mediator is studied in Ref.\,\cite{Archidiacono:2014nda}. The self-interacting sterile neutrino scenario has been tested against many cosmological observations \cite{Saviano:2014esa, Mirizzi:2014ama, Forastieri:2017oma, Choi:2018gho, Song:2018zyl}. It was shown in Ref.\,\cite{Chu:2018gxk} that the basic scenario is in conflict with bounds on the sum of neutrino mass, and some resolutions have been provided as well. One must note that these results depend on the choice of background cosmology as $\Lambda$CDM, and the bounds will change if other scenarios are considered \cite{RoyChoudhury:2018vnm,RoyChoudhury:2018gay,RoyChoudhury:2019hls}. Recent analyses of large-scale structure data have also indicated a preference for self-interacting neutrinos~\cite{He:2023oke,Camarena:2023cku}. With this as motivation, we ask if the dip due to a cosmic sterile-neutrino background can be probed in IceCube and whether the inclusion of these features leads to an appreciable effect on the inferred spectral index. The main result of this paper is shown in \Cref{fig:gamma}. We find that the systematic inclusion of the dips results in, 
$$ \gamma_{\texttt{\,astro}}^{\text{\,HESE,\,This Work}} = 2.60^{+0.20}_{-0.19} $$
which is in better agreement with the Northern Tracks sample. As discussed below, our errors are underestimated due to the limitations of our analysis. Using a simplified approach, we also show that the inclusion of these dips does not change the fits to the Northern Tracks significantly. However, for a quantitative assessment of the relaxation in the tension between the two datasets, a detailed Monte Carlo simulation, along with consistent treatment of backgrounds is required, which is beyond the scope of this work. We have shown in detail that an `SPL-with-dips' is a viable fit to the HESE sample and results in a harder spectral index than an SPL-only fit. Our conclusions are also applicable to other origins of the dip(s), for example, non-standard interactions of active neutrinos.

The paper is organized as follows. In \Cref{sec:methods}, we discuss the methodology followed in the paper. We discuss the model and constraints, the propagation of astrophysical neutrinos, our estimate of the event rates the three datasets (HESE, Northern Tracks, and ESTES), and our likelihood analysis. In \Cref{sec:results}, we provide the main results and end with a short summary in \Cref{sec:summary}.


\section{Model description and methods}\label{sec:methods}

\subsection{Lagrangian and constraints}
In this work, we follow the model of self-interacting sterile neutrino outlined in Ref. \cite{Dasgupta:2013zpn}. The sterile neutrinos are assumed to have self-interactions mediated by a new MeV-scale vector boson. The interaction Lagrangian is given by, 
\begin{equation}
	-\mathcal{L}_{\rm int}= g_X \bar{\nu}_s\gamma_\mu\frac{1}{2}(1-\gamma_5)\nu_sX^\mu\,,
	\label{eq:Lag}
\end{equation}
where $X^\mu$ is the mediator and $g_X$ is the self-interaction coupling constant. In this work, we fix the coupling strength as $g_X = 0.1$, which is sufficiently large for the re-coupling of sterile neutrinos \cite{Chu:2015ipa}.  At energy scales lower than the mediator mass ($M_X$), the interaction can be described by a Fermi-like contact interaction with an effective coupling given by $G_X = \frac{\sqrt{2}}{8} \frac{g_X^2}{M_X^2}$. Cosmological observations like big bang nucleosynthesis (BBN) and cosmic micorwave backgound (CMB) impose stringent constraints on this effective coupling~\cite{Song:2018zyl}. It was shown in Ref.~\cite{Song:2018zyl} that small values of the effective coupling do not sufficiently delay the production and are thus excluded. The minimum value of $G_X$ required to satisfy the BBN constraints at $95\%$ C.L. is $G_X \geq 10^4 G_F$. For the mediator mass range relevant to this work, i.e., $M_X\sim10-50$\,MeV, $g_X>0.01$ is required to satisfy the BBN constraints. We have shown resulting parameter space for both fixed-$g_X$ and fixed-$G_X$. 

The mass-eigenstates and the flavor-eigenstates are related by a $4\times4$ rotation matrix characterized by six mixing angles. The three mixing angles from the active sector are fixed to their measured values from Ref. \cite{Esteban:2020cvm}, and the remaining three angles are free parameters of the model. Due to this mixing, all four mass-eigenstates interact via the new mediator, and interactions between propagating mass-eigenstates and background neutrinos are possible \cite{Chauhan:2018dkd}. In this simplified model, we have five New Physics parameters -- three mixing angles ($\theta_{14}\,,\theta_{24}\,,\theta_{34}$), mass of the sterile neutrino ($m_s$), and mass of the mediator ($M_X$). In a UV-complete model, additional degrees of freedom may be required for generating the mixing and mass of the gauge boson, and for anomaly cancellation. 

\emph{Constraints from cosmology--} The self-interactions of the sterile neutrino delay their production and hence the constraints from BBN on $N_{\rm eff}$ can be avoided. However, the stringest constraints on the model come from the cosmological inference of the sum of neutrino masses \cite{Chu:2018gxk}. One of the possible ways to avoid this constraint is to assume additional light sterile neutrinos charged under the new gauge symmetry \cite{Chu:2018gxk}. In this scenario, when these sterile neutrinos recouple, the energy is redistributed among all new species which weakens the constraints on the sum of neutrino mass. In such a scenario, the number density of the cosmological eV-scale sterile neutrino background will be reduced by a factor proportional to the number of additional species. This can be compensated by a small increase in the coupling (as cross-section scales as $g_X^4$). This possibility was considered in Ref. \cite{Chauhan:2018dkd}. The other possibility is to consider a fast decay of the sterile neutrino to active neutrinos \cite{Chu:2018gxk}. As these sterile neutrinos decay after the decoupling, this would result in a new non-thermal distribution of active neutrinos ($f(p)\sim \delta(p - m_s/(2\,\text{or}\,3))$) and this may relax the neutrino mass bound\,\cite{Oldengott:2019lke}. The scattering of the cosmogeic neutrinos with the new component would result in similar ``dips", however this intersting possibility requires a separate and dedicated analysis.

\emph{Laboratory constraints--} In Ref. \cite{Fiorillo:2020zzj}, the constrains on self-interacting sterile neutrino from meson decays was evaluated assuming pseudo-scalar interactions. We expect that similar bounds will exist on vector mediators, however, we do not evlauate them. These bounds arise from measurements of leptonic decays of charged mesons ($M \rightarrow \ell + \nu_{\ell}$) which get additional contribution from self-interactions of the form $M \rightarrow \ell + \nu_{\ell} + X$ and $M \rightarrow \ell + \nu_{\ell} + \bar{\nu}_s + \nu_s$. The four-body decays are supressed, and dominant constraints arise from three-body decays only \cite{Fiorillo:2020zzj}. For the light sterile neutrino that is considered in this paper, the bounds on the coupling constant are strong when the new decay modes become kinematically accesible. Based on the results of Ref. \cite{Fiorillo:2020zzj}, we expect that our model is strongly constrained for $M_X<$\,2\,GeV, which is the entire parameter space in consideration. These bounds will be sensitive to details of UV completion, e.g., presence of additional light sterile neutrinos and other mediators. In this respect, the alternative model with a pseudoscalar mediator allows for selective flavor-dependence of the couplings (unlike gauge interactions that are universal) and the limits can be relaxed~\cite{Blinov:2019gcj}. Moreover, the pseudoscalar mediator is also cosmologically more viable too\cite{Dasgupta:2021ies}.

\subsection{Propagation of astrophysical neutrinos}
The introduction of this new interaction results in scattering between astrophysical neutrinos and the cosmic background of active and sterile neutrinos. Consequently, we expect dips and bumps in the flux of astrophysical neutrinos, which is otherwise assumed to have a single-power law behavior. The Boltzmann equation for the time evolution of the comoving number density of the astrophysical neutrinos per unit energy ($\tilde{n}_i = \partial \Phi_i/\partial E_\nu$) is given by \cite{Ng:2014pca}
\begin{align}
	\frac{\partial\tilde{n}_i(t,E_\nu)}{\partial t} & =  
	\frac{\partial}{\partial{E_\nu}} \left[H(t) \, E_\nu \, \tilde{n}_i(t,E_\nu)\right]+\mathcal{L}_i(t,E_\nu) \notag \label{eq:bolz} -\tilde{n}_i(t,E_\nu) \sum_j n^t_j \sigma_{ij}(E_\nu) \\
	& + \sum_{j, \, k, \, l} n^t_j  \int_{E_\nu}^\infty  dE_\nu' \, \tilde{n}_k(t,E_\nu') \frac{d\sigma_{jk \rightarrow il}}{dE_\nu}(E_\nu',E_\nu)\,.
\end{align}
In this equation, the first term on the right-hand side accounts for the cosmological expansion. The second term, called the \emph{source} term, accounts for the production of high-energy neutrinos due to astrophysical sources. In this work, we assume that the TeV--PeV neutrinos originate in $pp$-sources and following Ref. \cite{Esteban:2021tub}, the source term is taken to be a single power law with normalization proportional to the star-formation rate. The third term accounts for the interaction with background neutrinos which depends on the assumed number density of sterile neutrinos today. The sterile neutrinos are produced after recoupling from the population of active neutrinos. One can do detailed analysis by solving the 4-flavor Boltzmann transport equation (similar to Refs.~\cite{Song:2018zyl,Mazumdar:2019tbm}) to obtain the exact number density of the sterile neutrinos. However, we will assume flavor equilibrium for simplicity and assume that the number density of sterile neutrinos ($n_s$) can be approximated as $n_s = (1/4)\times n_a$, where $n_a = 2 \times 56 \,(1 + z)^3$ is the number density of active neutrinos per flavor in the standard scenario. The last term accounts for the \emph{regeneration} of lower-energy neutrinos after interactions. These equations have been solved using the publicly available code \text{nuSIprop} \cite{Esteban:2021tub}, which solves the Boltzmann equations for standard three-flavor neutrinos in the presence of non-standard interaction. We have extended the code to include the sterile neutrino. These are improvements on the work done in Ref. \cite{Chauhan:2018dkd} where the source term was ad-hoc, and regeneration was ignored.


\subsection{High Energy Starting Events (HESE)}
To estimate spectrum of HESE in IceCube, we use the Monte Carlo sample of simulated events\footnote{\url{https://github.com/icecube/HESE-7-year-data-release/}} from Ref. \cite{IceCube:2020wum}. The input parameters used to calculate the HESE event spectrum are tabulated in \Cref{tab:params}, and a description of these parameters can be found in Ref. \cite{IceCube:2020wum}. Of these, we only consider the astrophysical neutrino spectral index ($\gamma_{\texttt{astro}}$) and the normalization scale ($\Phi_{\texttt{astro}}$) as variables and fix other parameters to their typical values. We assume that the new physics does not affect the fluxes of conventional and prompt atmospheric neutrinos, as well as atmospheric muons. A detailed calculation where all parameters are optimized is computationally very expensive and beyond the scope of this work. The event spectrum and the subsequent log-likelihood are obtained by reweighting the Monte Carlo events from astrophysical neutrinos. As we also scan over the mixing angles, the reweighting accounts for the flavor dependence of the interactions. 

The IceCube collaboration uses the effective likelihood defined in Ref. \cite{Arguelles:2019izp} for estimating the physics parameters and their confidence and credible intervals \cite{IceCube:2020wum}. We follow their analysis closely, and present our results in the frequentist statistical methodology. For completeness, the likelihood analysis is summarised here and we refer the reader to Ref. \cite{IceCube:2020wum} and Ref. \cite{Arguelles:2019izp} for additional details. To handle the large MC statistical uncertainty due to small sample size in some bins, Ref. \cite{Arguelles:2019izp} introduces a modified Poisson likelihood function (called \emph{effective} likelihood), 
\begin{equation}
    \mathcal{L}_{\text{eff}}(\mu, \sigma, k) = \left( \dfrac{\mu}{\sigma^2}\right)^{\frac{\mu^2}{\sigma^2} + 1}\, \Gamma\left( k + \dfrac{\mu^2}{\sigma^2} + 1\right) \left[ k! \left(1 + \dfrac{\mu}{\sigma^2} \right)^{k + \frac{\mu^2}{\sigma^2} + 1} \Gamma\left(\dfrac{\mu^2}{\sigma^2} + 1\right)\right]^{-1},
\end{equation}
where $k$ is the observed number of events, $\mu$ is the expected number of events, $\sigma$ is the MC statistical uncertainty for the bin. The likelihood function is given by,
\begin{equation}\label{eq:lik}
    \mathcal{L}({\bm{\theta, \eta}}) = \prod_{j}^{N_{\text{bins}}} \, \mathcal{L}_{\text{eff}}( \mu_j({\bm{\theta, \eta}}), \sigma_j({\bm{\theta, \eta}}), k_j) 
\end{equation}
where $N_{\text{bins}}$ is the number of bins, $\bm{\theta}$ is the set of parameters of interest, and $\bm{\eta}$ is the set of nuisance parameters. Note that, for a given analysis the set of nuisance parameters also includes physical parameters that are not being examined. In Ref. \cite{IceCube:2020wum}, the data and the MC events are first separated by their morphology (tracks, cascades, double-cascades) and then binned further in two observables. The tracks and cascades are binned in their reconstructed energy and reconstructed zenith angle, whereas the double cascades are binned in reconstructed energy and their reconstructed separation between cascades. The bin widths are chosen to be comparable to the detector resolution and there are 840 bins in observable quantities. These bins contain contributions from atmospheric as well as astrophysical events. 

\begin{table}[b]
         \caption{ \label{tab:params} The set of physical and nuisance parameters that determine the HESE spectrum are tabulated. We have shown the best-fit parameters for single power law (SPL) obtained in Ref.\,\cite{IceCube:2020wum} and the values/range of parameters used in this work. For a description of these variables, see TABLE VI.1 in Ref.\,\cite{IceCube:2020wum}}
	\centering
	\begin{tabular}{c c c}
		\toprule[.75pt]
		\hline
		\textbf{Parameter} & \quad\textbf{SPL best-fit} & \quad\textbf{This Work} \\
		\hline
		\midrule
		$\gamma_{\texttt{astro}}$ & 2.874 & (2,4) \\
		$\Phi_{\texttt{astro}}$ & 6.365 & (1,10) \\ 
		$\Phi_{\texttt{conv}}$ & 1.006& 1.000 \\ 
		$\Phi_{\texttt{prompt}}$ & 0.0 & 0.0 \\ 	
	    $\Phi_{\texttt{muon}}$ & 1.187 & 1.00 \\ 
		\hline
		$\Delta \gamma$ & -0.053 & -0.050 \\ 
		$\nu /\overline{\nu}$ & 0.998 & 1.000 \\ 
		$a_s$ & 1.001 & 1.000 \\ 
		$\varepsilon_{\rm dom}$ & 0.952 & 0.990 \\ 
		$\varepsilon_{\rm h.o.}$ & -0.055 & 0.000 \\ 
		$K/\pi$ & 1.000 & 1.000\\
		\toprule[.5pt]
		\hline
	\end{tabular}
\end{table}

We follow the analysis of Ref.\,\cite{IceCube:2020wum} closely to estimate the new physics parameters that are most likely to explain the \text{HESE} data, and compare it with the single power-law. The \emph{dips} in the astrophysical neutrino spectrum will not affect the events from atmospheric neutrinos and muons. Moreover, the atmospheric contribution to the starting events in IceCube is primarily determined be the low-energy events and the zenith-angle distributions. Hence, in our analysis we will fix the parameters associated with atmospheric events to their nomianal values\footnote{We have verified that keeping the atmospheric parameters free does not change the results significantly. This is primarily because the atmospheric parameters are determined by the lower energy events and their zenith-angle distributions.}, and only analyse astrophysical parameters (i.e., $\gamma_{\texttt{astro}}$ and $\Phi_{\texttt{astro}}$). Due to this approximation, the uncertainities are underestimated and a comparison in presented in Appendix A.  The best-fit parameters and the confidence intervals are obtained using the profile likelihood, 
\begin{equation}\label{eq:lik_pro}
    \tilde{\mathcal{L}}^{\texttt{profile}}(\bm{\theta}) = \max_{\bm{\eta}} \mathcal{L}(\bm{\theta, \eta})\cdot \Pi(\bm{\theta, \eta})
\end{equation}
where $\Pi(\bm{\theta, \eta})$ denote the constraints on parameters. For parameters for which no external data is available, $\Pi$ is constant. We have used the Python implementation of \text{SLSQP} algorithm to minimize the negative of the log of the function and thus obtain the best-fit point $\hat{\bm{\theta}}$. The model parameter test statistic is defined as, 
\begin{equation}\label{eq:ts}
    \text{TS}(\bm{\theta}) = - 2\,\log\,\left( \dfrac{\tilde{\mathcal{L}}^{\texttt{profile}}(\bm{\theta})}{\tilde{\mathcal{L}}^{\texttt{profile}}(\hat{\bm{\theta}})} \right).
\end{equation}
The results are presented using Wilks' theorem with appropriate degrees of freedom \cite{Wilks:1938dza}. 

\subsection{Glashow Resonance}

The HESE sample only includes events for which the interaction vertex is inside the restricted fiducial volume. To enhance the sensitivity to Glashow resonance (GR) events, a larger fiducial volume is required, which is achieved by looking at PeV energy partially contained events (\text{PEPE}). The Monte Carlo simulation provided with the HESE data release can be utilised to predict the number of GR events in the PEPE sample by a simple rescaling. Due to the larger effective area, the expected event rate for the PEPE selection is about twice that of the HESE selection \cite{IceCube:2021rpz}. However, the livetime of PEPE sample is only 4.6 years as opposed to HESE for which the livetime is 7.22 years. If $\lambda_{\text{HESE}}$ is the event rate in the GR energy bin (i.e., 5.6\,PeV to 7.7\,PeV ), obtained from the simulation, then one can estimate the expected number of events in the PEPE sample as, 
\begin{equation}
    \lambda_{\text{PEPE}}(\bm{\theta, \eta}) = \lambda_{\text{HESE}}(\bm{\theta, \eta}) \times 2 \times (4.6/7.22)
\end{equation}
We have verified that the estimates obtained using this method agree very well with the Extended Data Table 1 in Ref. \cite{IceCube:2021rpz}. As only one event has been observed in this bin, we define the likelihood assuming Poisson statistics as,
\begin{equation}
    \mathcal{L}_{\text{GR}}(\bm{\theta, \eta}) \approx \lambda_{\text{PEPE}}(\bm{\theta, \eta})\, e^{- \lambda_{\text{PEPE}}(\bm{\theta, \eta})}.
\end{equation}
The likelihood function, i.e., $\mathcal{L}(\bm{\theta, \eta})$ in \Cref{eq:lik}, is multiplied by $\mathcal{L}_{\text{GR}}(\bm{\theta, \eta})$ to obtain the combined fit. The impact of including the GR event in our analysis is shown in Appendix A. 

\subsection{Northern Tracks}\label{Sec:Methods-NT}

The Northern Tracks analysis by IceCube focusses on measuring the flux of high-energy $\nu_\mu$ and $\bar{\nu}_\mu$ from atmospheric and astrophysical sources. The event selection criterion samples muon-like\footnote{We refer to both $\mu^-$ and $\mu^+$ as ``muons"} tracks originating from the Northern celestial hemisphere, which significantly reduces the background of atmospheric muons and results in high-purity neutrino-induced events. The sample contains mostly through-going muon tracks and a small contribution from starting tracks. Unlike the HESE sample, a detailed MC simulation and likelihood analysis for Northern Tracks is not publicly available. In order to estimate the impact of dips in the neutrino spectrum on Northern Tracks, we use a ``theorist's approach" \cite{Kistler:2006hp, Laha:2013lka, Gaisser:2016uoy} to calculate the event rates.

The measurable muon spectrum for Northern Tracks is given by, 
\begin{equation}
    \dfrac{dN_{\text{NT}}}{dE_\mu^{\text{true}}} = \dfrac{dN_{\text{thru. tr.}}}{dE_\mu^{\text{true}}} + \dfrac{dN_{\text{st. tr.}}}{dE_\mu^{\text{true}}}  
\end{equation}
where the first term is the contribution from the through-going tracks that originate outside the detector volume, and the second term denotes the contribution of the starting tracks for which the primary interaction vertex is inside the detector volume. We have used $E_\mu^{\text{true}}$ to denote the true energy of the muon which is different from the reconstructed muon energy. The muon spectrum from the through-going events is calculated using, 
\begin{equation}\label{eq:full_NT}
    \dfrac{dN_{\text{thru. tr.}}}{dE_\mu^{\text{true}}} = \sum_{\nu_\mu, \bar{\nu}_\mu}\dfrac{T\,N_A}{\alpha + \beta E_\mu^{\text{true}}} \int d \Omega \int_{E_\mu^{\text{true}}}^{E_\nu^{\text{max}}(\theta_z)} dE_\nu\, A(E_\nu, \theta) e^{-\uptau(E_\nu, \theta_z)}\,\dfrac{d\Phi_\nu}{dE_\nu}  \int_{0}^{1 - E_\mu^{\text{true}}/E_\nu} dy\, \dfrac{d\sigma_\nu (E_\nu, y)}{dy},
\end{equation}
where $d\sigma/dy$ is the differential neutrino-nucleon deep-inelastic scattering cross-section with respect to the inelasticity parameter $y = E_\mu/E_\nu$, $\uptau$ is the optical depth that accounts for the interaction of neutrinos as they propagate inside earth \cite{Gandhi:1995tf}, $A$ is the energy and zenith-angle dependent effective area of the detector, $\alpha$ and $\beta$ are the related to the stochastic energy-loss of muons\footnote{$\alpha\approx 2\times10^{-3}$\,GeV\,cm$^2$/g, $\beta\approx 5\times 10^{-6}$\,cm$^2$/g, \cite{Koehne:2013gpa}}, $T$ is the livetime of the detector, and $N_A$ is the Avogadro's constant. In \Cref{eq:full_NT}, $E_\nu^{\text{max}}$ is the maximum value of the initial neutrino energy that results in a muon entering the fiducial detector volume with energy $E_\mu^{\text{true}}$ and ideally depends on the zenith-angle $\theta_z$ \cite{Laha:2013lka}. One must note that the neutrino flux in \Cref{eq:full_NT}, i.e., $d\Phi_\nu/dE_\nu$, is for a single neutrino species. Lastly, there is a sum over $\nu_\mu$ and $\bar{\nu}_\mu$ as the differential cross-section and the total cross-section (which enters the expression via $\uptau$) are different for neutrinos and anti-neutrinos. We have used the \texttt{NNPDF40} parton distribution functions \cite{NNPDF:2021njg} evaluated with \texttt{LHAPDF} \cite{Buckley:2014ana} to calculate the differential and total cross-sections assuming a isoscalar target nucleon \cite{Gandhi:1995tf}. The optical depth $\uptau$ has been evaluated using the preliminary reference earth model (PREM) density profile \cite{Dziewonski:1981xy}. 

The evaluation of \Cref{eq:full_NT} is not only computationally expensive but also require input from IceCube about effective areas. We can simplify the expression by first assuming that $A(E_\nu, \theta) \approx A = 1\,\text{km}^2$. We also assume that $E_\nu^{\text{max}}(\theta_z) \rightarrow \infty$. As a result, the expression can be simplified as, 
\begin{equation}\label{eq:approx_NT}
    \dfrac{dN_{\text{thru. tr.}}}{dE_\mu^{\text{true}}} = \dfrac{2\pi\,T\,N_A\,A}{\alpha + \beta E_\mu^{\text{true}}} \int_{E_\mu^{\text{true}}}^{\infty} dE_\nu\, \mathcal{S}_{\text{NT}}(E_\nu)\,  \tilde{\sigma}( E_\mu^{\text{true}} , E_\nu)\,\dfrac{d\Phi_{\nu + \bar{\nu}}}{dE_\nu}
\end{equation}
where, 
\begin{equation}
    \tilde{\sigma}( E_\mu^{\text{true}} , E_\nu) = \int_{0}^{1 - E_\mu^{\text{true}}/E_\nu} dy\, \dfrac{1}{2} \,\left[\dfrac{d\sigma_\nu (E_\nu, y)}{dy} + \dfrac{d\sigma_{\bar{\nu}} (E_\nu, y)}{dy} \right]
\end{equation}
and the average attenuation (or \emph{shadow}) factor for Northern Tracks is,
\begin{equation}
    \mathcal{S}_{\text{NT}}(E_\nu) = \int_{-1}^{0} d(\cos \theta_z) \, \dfrac{1}{2}\, \left[ e^{-\uptau(E_\nu, \theta_z)} + e^{-\uptau(E_{\bar{\nu}}, \theta_z)}\right].
\end{equation}
Note that in \Cref{eq:approx_NT}, the flux accounts for \emph{both} $\nu_\mu$ and $\bar{\nu}_\mu$. We have neglected the contributions from leptonic interaction channels such as $\bar{\nu}_e+e^- \rightarrow \bar{\nu}_\mu + \mu^-$ as their cross-sections are significanly smaller than DIS.

For the starting-tracks contribution, we have followed Ref. \cite{Mena:2014sja} and use the same approximations to obtain, 
\begin{equation}\label{eq:NTST}
    \dfrac{dN_{\text{st. tr.}}}{dE_\mu^{\text{true}}} = 2\pi\,T\,N_A\, \rho\, V \int_{0}^{\infty} dE_\nu \,\mathcal{S}_{\text{NT}}(E_\nu) \,\dfrac{d\sigma(E_\mu, E_\nu)}{dE_{\mu}}\,\dfrac{d\Phi_{\nu + \bar{\nu}}}{dE_\nu} 
\end{equation}
where the differential cross-section $d\sigma/ d E_\mu$ is averaged over neutrino and anti-neutrino, $\rho=0.9167$\,g/cm$^3$ is the density of ice, and we take $V=1\,\text{km}^3$. 

Lastly, we need to account for the detector resolution. The true energy deposited by the travelling muon ($E_\mu^{\text{true}}$) and the proxy muon energy that is reconstructed do not usually have a one-to-one mapping. We model this uncertainty as a Gaussian, and the measured muon spectrum is given by \cite{Mena:2014sja},   
\begin{equation}\label{eq:Gauss}
    \dfrac{dN}{dE_\mu} = \int_{0}^{\infty} d E_\mu^{\text{true}} \, \dfrac{dN}{dE_{\mu}^{\text{true}}}\, \mathcal{G}\left( E_\mu, E_\mu^{\text{true}}, \sigma(E_\mu^{\text{true}})\right)
\end{equation}
where we have considered $\sigma(E_\mu^{\text{true}}) = 0.125\,E_\mu^{\text{true}}$ in our analysis, similar to Ref. \cite{Mena:2014sja}. We understand that the starting track events have a better energy resolution than the through-going tracks, however, we do not make this difference. We use \Cref{eq:approx_NT} and \Cref{eq:Gauss} to estimate the contribution of astrophysical neutrinos to the Northern Tracks. For atmospheric contribution, we use the results in Ref. \cite{IceCube:2025ary}. 

The observed data for Northern tracks is provided in Ref.\cite{Abbasi:2021qfz} , which we use for our likelihood analysis. The IceCube collaboration uses both energy distribution and zenith-angle distribution of the observed events to calculate the likelihood. The atmospheric neutrino contribution to the Northern Tracks dominates at lower energies and has a very strong and characteristic zenith-angle dependence. In this work, we only consider the muon-energy bins above 100 TeV to minimize the effect of atmospheric neutrinos. We do not consider binning in zenith-angle as the atrophysical neutrinos have almost flat distribution. We define the likelihood assuming Poisson distribution as, 
\begin{equation}\label{eq:lik_nt_bin}
    \mathcal{L}_i(\bm{\theta} , \bm{\eta}) = \frac{\lambda_i(\bm{\theta}, \bm{\eta})^k e^{-\lambda_i(\bm{\theta}, \bm{\eta})}}{k!}
\end{equation}
where $\lambda_i(\bm{\theta}, \bm{\eta})$ is the expected number of events in the $i$-th bin obtained by integrating \Cref{eq:Gauss}, and $k$ is the number of number of observed events in the bin. In our analysis, we only consider the bins for which events are recorded and we do not consider the zenith-angle distribution of events. The likelihood, profile likelihood, and the test statistic are defined similar to \Cref{eq:lik}, \Cref{eq:lik_pro}, and \Cref{eq:ts}, respectively.

\subsection{Enhanced Starting-Track Event Selection (ESTES)}\label{Sec:Methods-ESTES}

The Enhanced Starting-Track Event Selection (ESTES) in IceCube is a event selection criterion that focuses on identifying $\nu_\mu$ and $\bar{\nu}_\mu$ interactions that originate inside the detector volume from all directions (northern sky and southern sky) \cite{Silva:2021kvx,Silva:2019fnq,Mancina:2019hsp,Mancina:2021jbk,IceCube:2021ctg,Silva:2023wol,Silva:2023mrh,IceCube:2024fxo}. By employing a dynamical veto volume and neutrino self-veto, the backgrounds from atmospheric muons and atmospheric neutrinos can be efficiently rejected, especially from the southern sky. This allows IceCube to probe for astrophysical muon neutrinos down to 1 TeV. ESTES events benefit from good directional resolution and improved neutrino energy reconstruction compared to other event selections, and have enchanced sensitivity to southern-sky neutrino sources, including the Galactic plane. These event can also be used to study the diffuse flux of neutrinos. Assuming a single power-law, the best fit spectral index for astrophysical neutrinos is, 
\begin{equation}
    \gamma^{\text{ESTES}}_{\texttt{astro}} = 2.58^{+0.10}_{-0.09},
\end{equation}
which lies between the fits obtained from HESE and Northern Tracks. 

In order to estimate the impact of dips on the ESTES data, we again adopt a \emph{theorist's approach} to event rates. The muon spectrum for starting-tracks can be obtained using \Cref{eq:NTST} and \Cref{eq:Gauss}. As the ESTES sample looks at events from all directions, the shadow factor can be approximated as,
\begin{equation}
    \mathcal{S}_{\text{ESTES}}(E_\nu) = 1 + \mathcal{S}_{\text{NT}}(E_\nu),
\end{equation}
assuming that the events from the southern-sky are not attenuated. Our approximation for the energy resolution should work for ESTES data as well\cite{IceCube:2024fxo}.  

Similar to Northern Tracks, the IceCube collaboration uses both muon-energy and zenith-angle distribution for their statistical analysis. In this work, we will only only consider the muon-energy bins above 70 TeV for the likelihood analysis of the ESTES data. The bin likelihood, total likelihood, profile likelihood, and the test statistic are defined similar to \Cref{eq:lik_nt_bin}, \Cref{eq:lik}, \Cref{eq:lik_pro}, and \Cref{eq:ts}, respectively.


\section{Searching for dips in spectrum}\label{sec:results}

\subsection{HESE and GR Analysis}

In this section, we look at the range of new physics parameters that lead to significant dips in the HESE sample and evaluate their impact on the spectral index ($\gamma_{\texttt{astro}}$) and the normalization scale ($\Phi_{\texttt{astro}}$) of the diffuse astrophysical flux. In our analysis, we fix the values of the parameters related to the detector efficiency, cosmic ray flux, and atmospheric neutrinos (see \Cref{tab:params}). As a result of this approximation, the errors in this paper are underestimated. A detailed scan of all parameters is computationally intensive and beyond the scope of this work. The parameters of interest are, 
\begin{equation}
    \bm{\theta} = \{ \gamma_{\texttt{astro}},\, \Phi_{\texttt{astro}},\, M_X,\, m_s,\, \theta_{14},\, \theta_{24},\, \theta_{34} \}.
\end{equation}
The best-fit values of these parameters from our analysis is tabulated in \Cref{tab:bf}. We fix the mixing angles to their best-fit values for the rest of the analysis. This reduces the computation time significantly and allows us to scan over relevant parameter space only. This approximation underestimates the 90\% C.L. region, and the 68 \% C.L. region is mostly unaffected. 

\begin{figure}[t]
    \centering
    \includegraphics[width=0.45\linewidth]{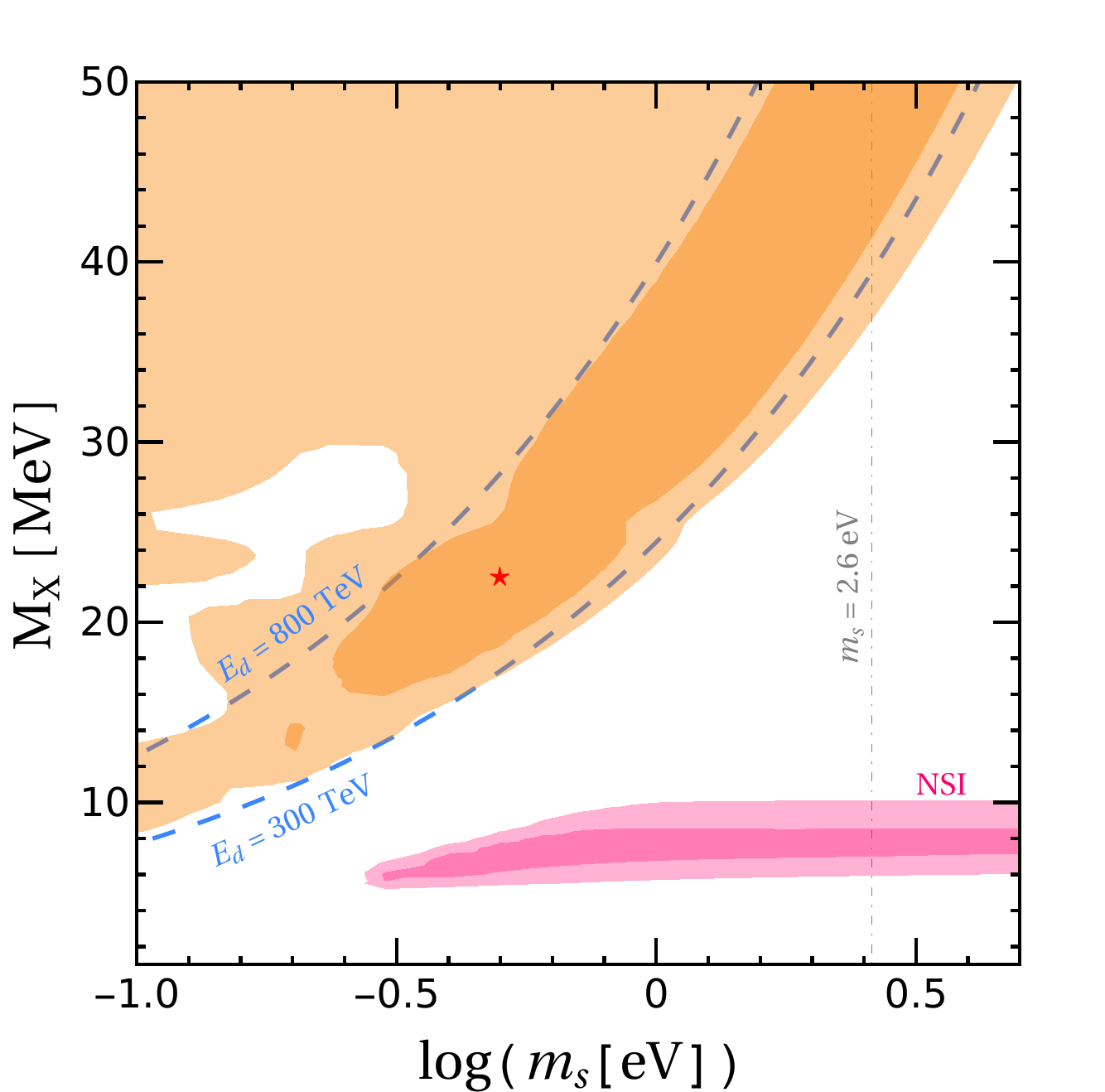}
    \includegraphics[width=0.45\linewidth]{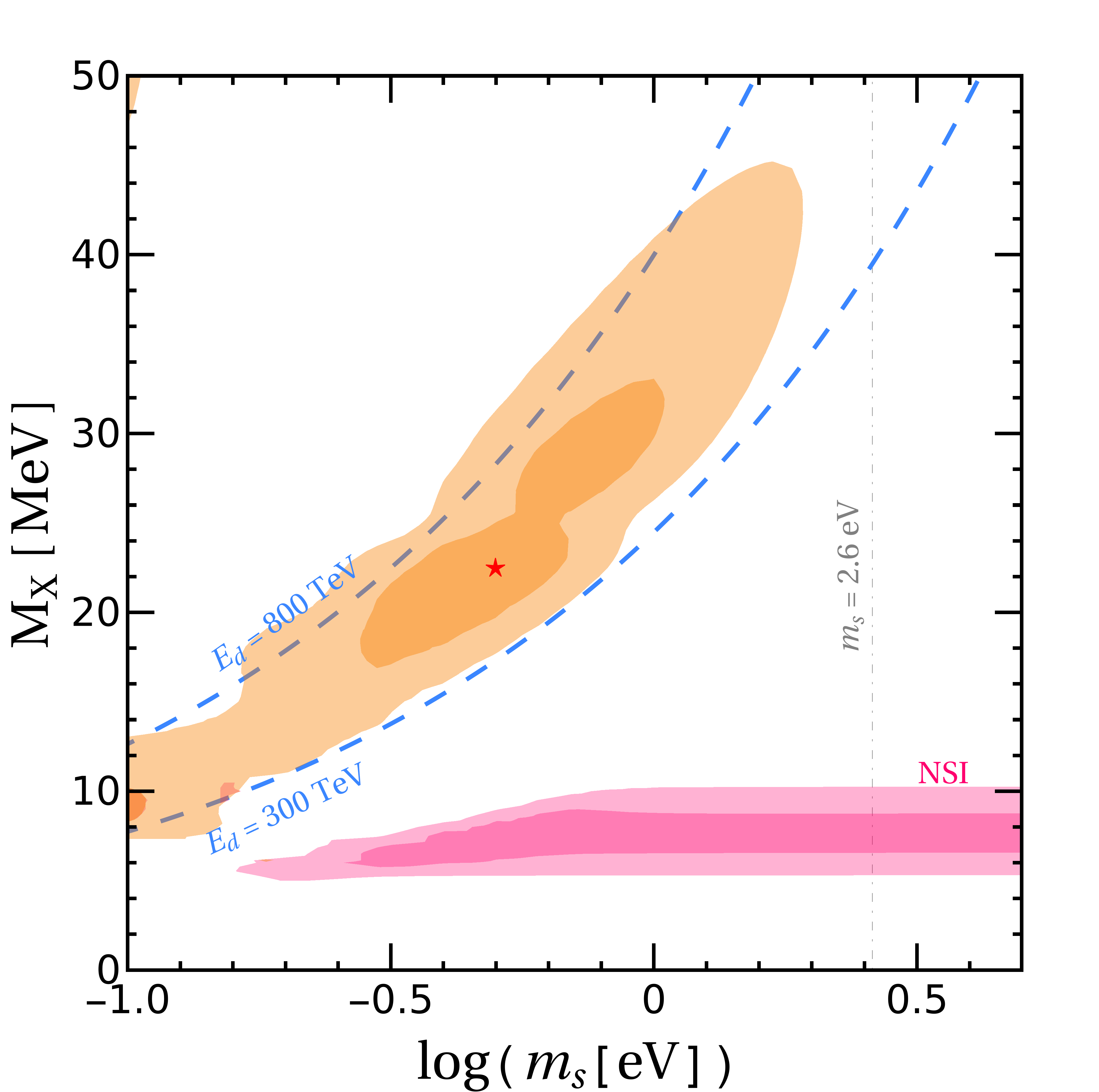}
    \caption{\emph{Left:} Scan of the profile likelihood (see \Cref{eq:ts}) for the two new physics parameters: mass of the sterile neutrino ($m_s$) and mass of the vector-boson mediator ($M_X$), is shown in orange. The darker (lighter) region denotes the 68\% C.L. (90 \% C.L.) interval. Note that for each point, the mixing angles are fixed to their best-fit values, and all other parameters are optimized. In this plot, we have fixed $g_X=0.1$. We also show two approximate curves that result in dips at 300 TeV and 800 TeV due to interaction with a cosmic sterile neutrino background. The region labeled as NSI (pink) indicated the parameter space for which dips occur due to interaction with the active neutrino background only and the dip due to sterile neutrino is outside the HESE analysis bound. \emph{Right:} Same as the left figure but for $G_X = 10^6\,G_F$ }
    \label{fig:MXms}
\end{figure}

\begin{table}[b]
	\caption{ \label{tab:bf} The best-fit values of the parameters of interest for single power law (SPL) obtained in Ref.\,\cite{IceCube:2020wum} and the result from our HESE+GR analysis is tabulated.}
	\centering
	\begin{tabular}{c c  c}
            \toprule[.75pt]
		\textbf{Parameter} & \quad\textbf{SPL best-fit\quad} & \quad\textbf{This Work\quad} \\
		\hline
		\midrule
		$\gamma_{\texttt{astro}}$ & 2.874 & 2.613 \\
		$\Phi_{\texttt{astro}}$ & 6.365 & 5.461 \\ 
		$M_X$ & -- & 22.96 MeV \\ 
		$m_s$ & -- & 0.50 eV \\ 
		$\theta_{14}$ & -- & 0.52 \\ 
		$\theta_{24}$ & -- & 0.53 \\ 
		$\theta_{34}$ & -- & 0.41 \\
		\bottomrule[0.5pt]
		\hline
	\end{tabular}

\end{table}

We first look at the space of the new physics parameters -- sterile neutrino mass ($m_s$) and the vector mediator mass ($M_X$), and present our results in \Cref{fig:MXms}. As expected\footnote{The \Cref{fig:MXms} should be compared with Fig. 2 in \cite{Chauhan:2018dkd}}, we find a large degenerate parameter space that results in dips in the event spectrum. We also find an island for $M_X=5-10$ MeV, which is largely independent of the sterile neutrino mass. In this region, the dip due to sterile neutrino is below the 60 TeV threshold of the HESE sample, and there is a dip between 300-800 TeV due to the heavier active neutrino. This is similar to the case of neutrino secret interaction ($\nu$SI), which is considered in Ref. \cite{Ng:2014pca}. We also highlight the $m_s = 2.6$ eV that is hinted in recent results \cite{Serebrov:2020kmd, IceCubeCollaboration:2022tso}.

\begin{figure}[h!]
	\centering
        \includegraphics[width=0.5\linewidth]{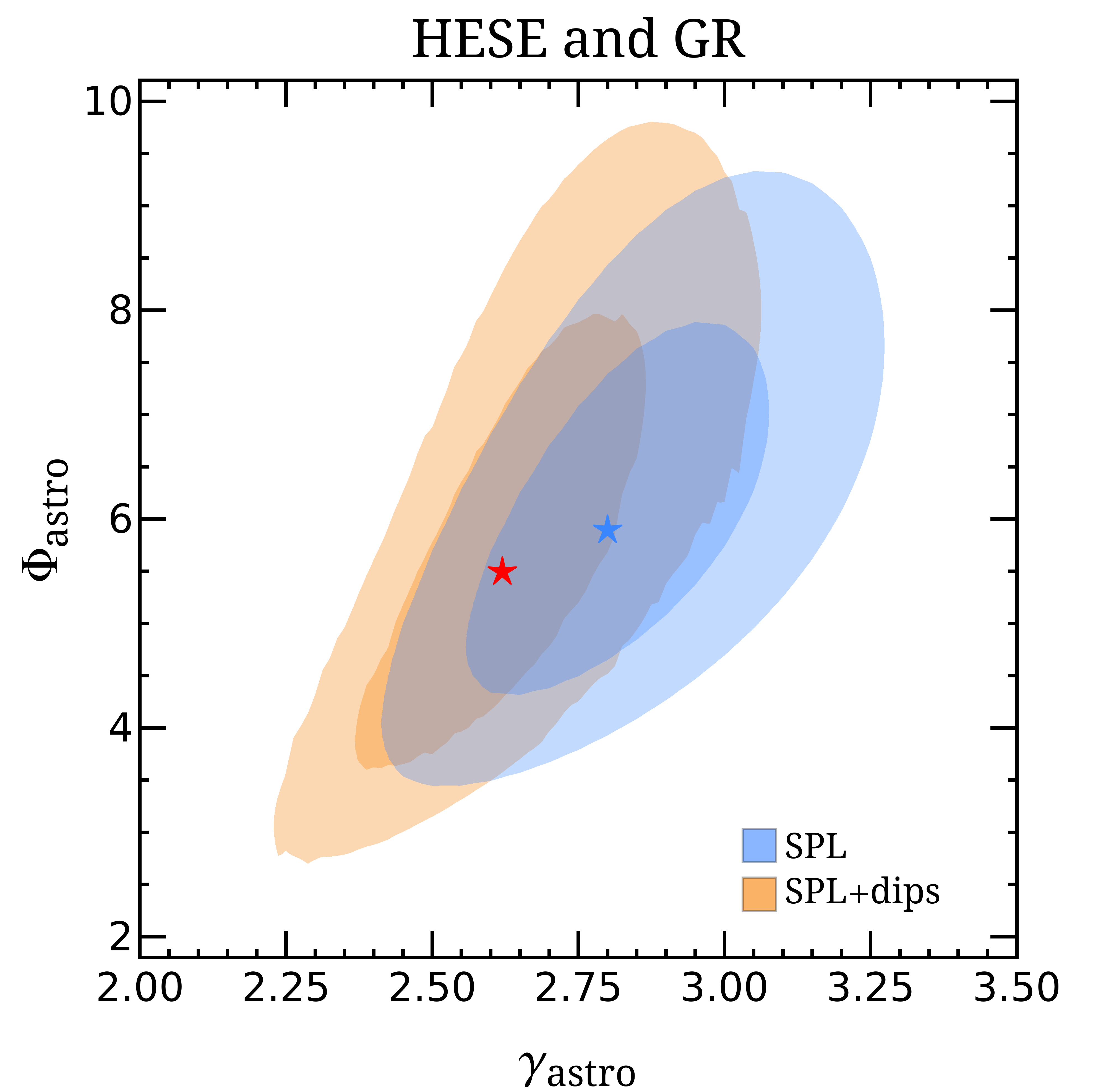}
	\caption{\label{fig_phigamma} Scan of the profile likelihood for the two astrophysical neutrino flux parameters: normalization scale and spectral index obtained from analysis of HESE and GR data. The darker (lighter) regions show the 68 \% C.L. (90 \% C.L.) intervals. The region in blue is obtained using single power law, and the region in orange is obtained by including the dips due to sterile neutrino background. Note that other physical parameters are optimized for each point. As we do not optimize over all of the parameters (see text for details), the intervals are slightly smaller than the ones reported in Ref.\,\cite{IceCube:2020wum}. See Appendix A for comparison. }
\end{figure}

We now look at the space of astrophysical neutrino flux parameters -- spectral index ($\gamma_{\texttt{astro}}$) and the normalization scale ($\Phi_{\texttt{astro}}$). The scan of the profile likelihood is presented in \Cref{fig_phigamma}. We find that the inclusion of the dips leads to lower values of the spectral index. The inclusion of the Glashow resonance event also disfavors large values of the spectral index. This is one of the key results of the paper. At present, with only 7.5 years of data, there is a large overlap between the preferred regions for SPL and SPL-with-dips. However, the trend is clearly noticeable. In the future, with additional data from other neutrino telescopes, we anticipate that the overlap will be resolved. This was recently demonstrated in Ref. \cite{Fiorillo:2022rft}, where the authors search for a bump-like feature in the spectrum. We also look at single parameter profile likelihood, and the results have been presented in \Cref{fig:gamma}. It is clear that due to the inclusion of dips in the spectrum, the fits to HESE and GR data prefer a harder spectral index. 

\begin{figure}[h!]
	\centering
        \includegraphics[width=0.49\linewidth]{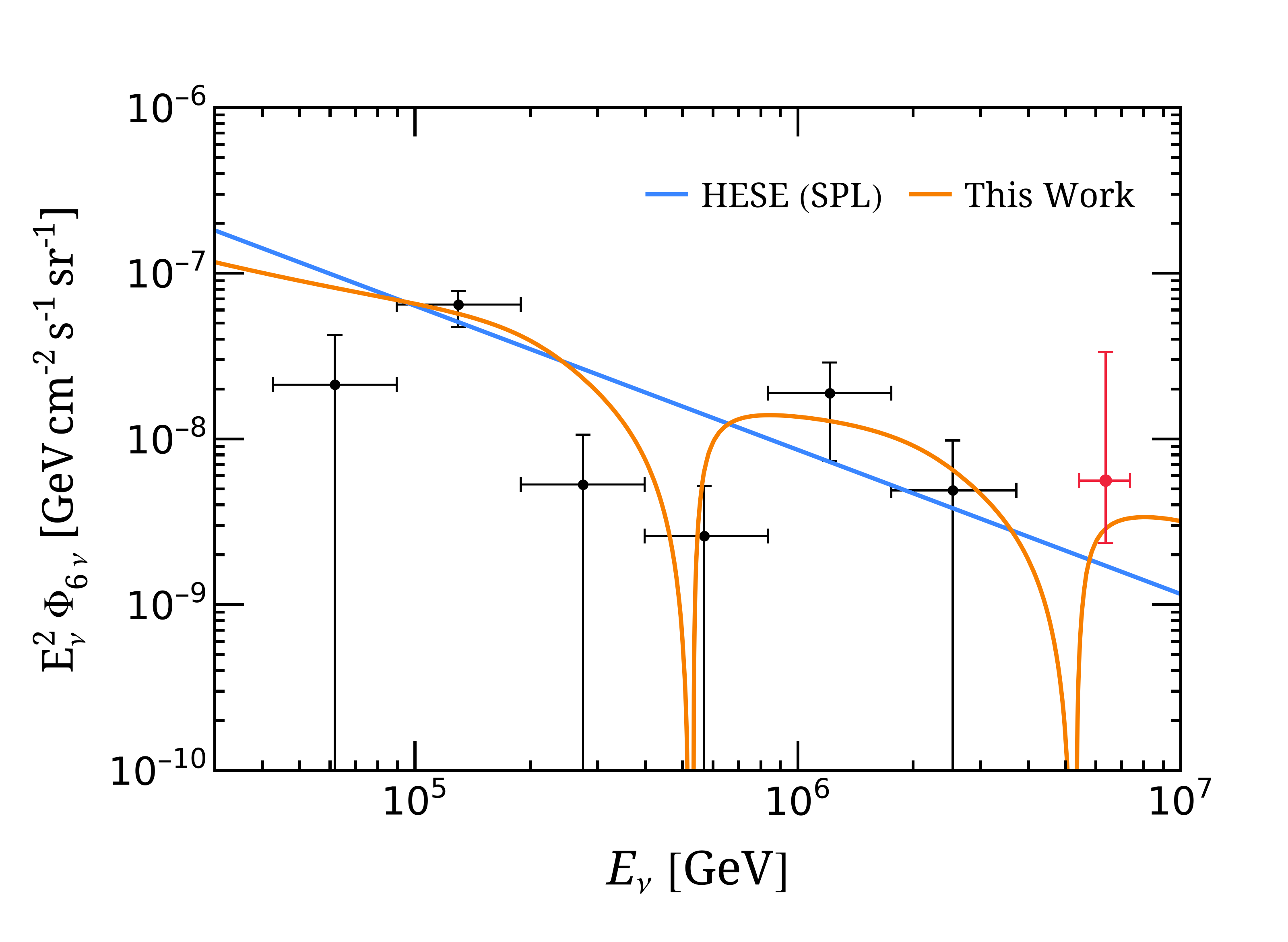}
	\includegraphics[width=0.47\linewidth]{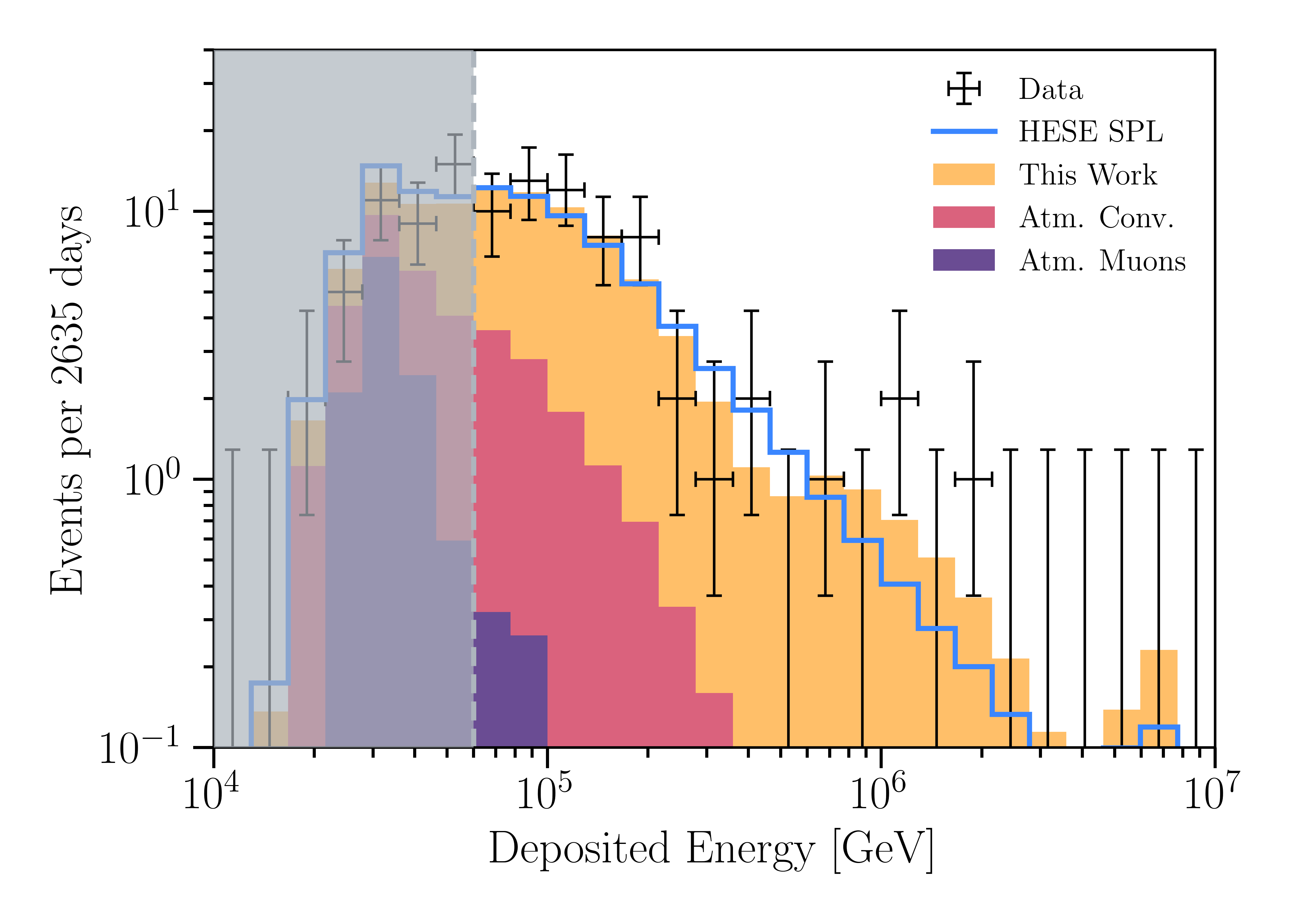}
	\caption{\label{fig:flux_events} \emph{Left:} We show the best-fit SPL flux from Ref. \cite{IceCube:2020wum} (blue) and the best-fit flux from our analysis (orange). The binned reconstructed flux from the HESE sample \cite{IceCube:2020wum} and PEPE \cite{IceCube:2021rpz} sample is shown in black and red, respectively. \emph{Right:} The deposited energy spectrum of HESE events (black) is shown with our event prediction using the best-fit parameters for SPL with dips (orange) and SPL-only (blue). The background from atmospheric neutrinos and muons is the same for both curves. The shaded grey region represents the 60 TeV threshold below which the backgrounds dominate the event rate.
 }
\end{figure}

In order to highlight the difference between the SPL best-fit and the best-fit parameters for SPL-with-dips, we show the neutrino flux prediction overlaid with the reconstructed flux from the IceCube collaboration in \Cref{fig:flux_events}. This plot only indicates of the position of dips in the neutrino spectrum. In our analysis, the likelihood is computed using the data from the deposited energy event spectrum. We have shown the predicted event spectrum from the two sets of best-fit parameters in \Cref{fig:flux_events}. As mentioned earlier, we have verified that the atmospheric neutrino paramters are unaffected by the new physics, and the background predictions for both fits are statistically indistinguishable. In a more detailed analysis, one may consider varying the models for atmospheric neutrinos. As there is a large overlap in the 68\% C.L. interval of the two fits, we choose not to show the uncertainty bands in these figures.  

\subsection{Northern Tracks Analysis}

In \Cref{Sec:Methods-NT}, we outlined our methodology for estimating the distribution of Northern Track events in IceCube. We now compare our predictions with observed IceCube data to extract the astrophysical flux normalization and spectral index. We find that our methods systematically underestimate the event rates. This is a significant limitation of our analysis, and our best-fit normalization differs substantially from the official IceCube results. A further limitation of our analysis is the treatment of the atmospheric backgrounds. The fit of atmospheric neutrino parameters are determined by the low-energy data as well as their zenith-angle distribution, and is largely independent of the dips. Subsequently, we adopt the atmospheric component as reported by IceCube, and only use the high-energy bins ($>100\,$TeV) for our likelihood analysis. Moreover, we do not utilise the zenith-angle distribution in our analysis. As a result of this simplified treatment of statistics, the uncertainities on our analysis are over-estimated. Therefore, our results should be interpreted in terms of trends and relative comparisons rather than as precise determinations of the flux parameters.

\begin{figure}[h!]
    \centering
    \includegraphics[width=0.40\linewidth]{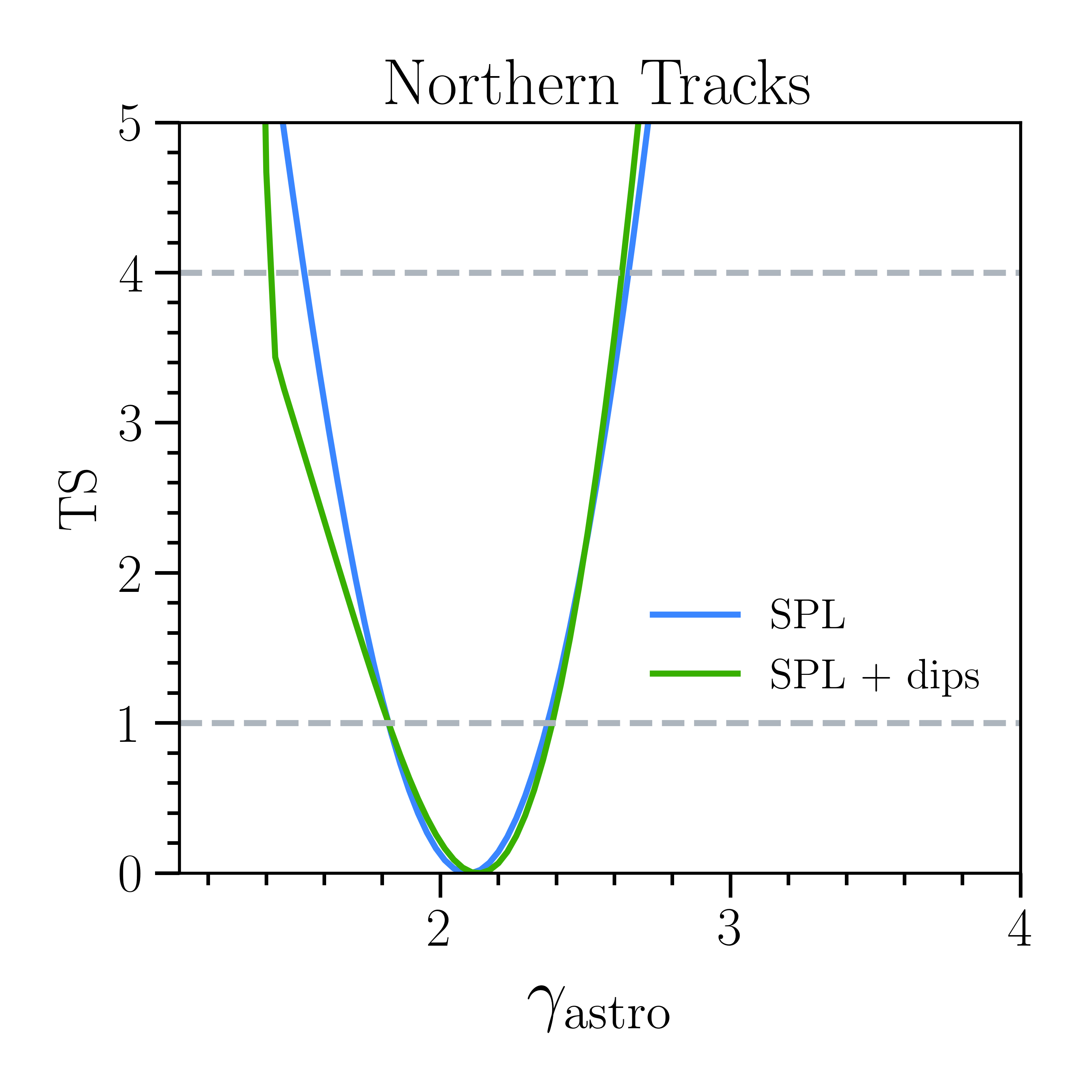}
    \includegraphics[width=0.56\linewidth]{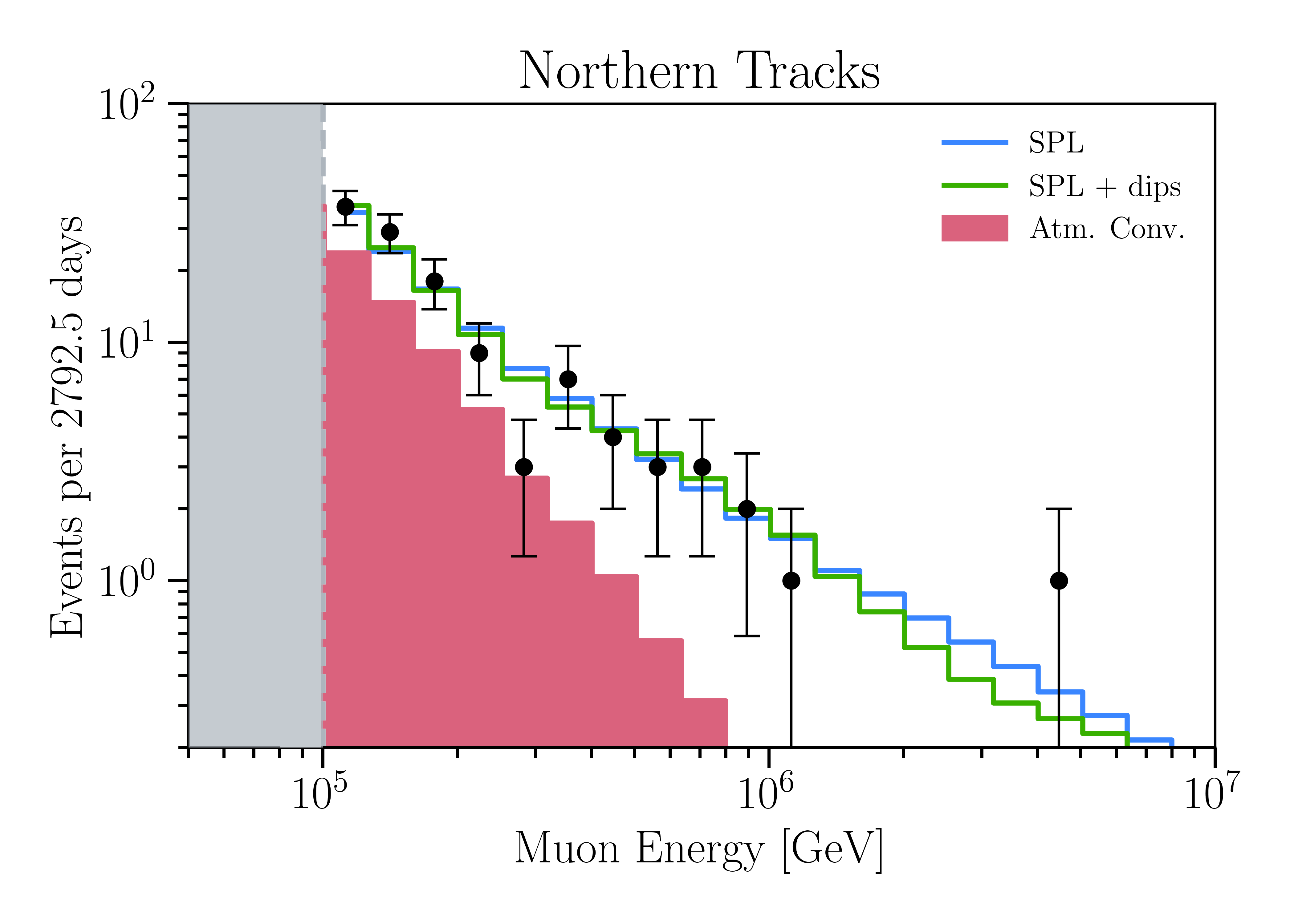}
    \caption{
    \emph{Left:} The test statistic (see \Cref{eq:ts}) for the single power-law fit of 9.5 years Northern Tracks sample \cite{Abbasi:2021qfz} is shown in blue. In green, we show the test statistic after including absorption features using the best-fit New Physics parameters from \Cref{tab:bf}. The horizontal grey dashed lines are used to denote the 68\% C.L. and 90\% C.L. intervals. \emph{Right:} The event rates obtained using \Cref{eq:Gauss} for the SPL and SPL+dips are shown in blue and green respectively. Note that the normalization for both fits is slightly different. The observed data points are taken from \cite{Abbasi:2021qfz} and the errors are statistical only. The red shaded region is the atmospheric contribution. In the grey shaded region, the atmospheric contribution is large and hence excluded from our analysis (see text for details).
    }
    \label{fig:NT}
\end{figure}

With these limitations in mind, we examine whether the Northern Track data can discriminate between two representative flux hypotheses, i.e., SPL and SPL-with-dips. For the SPL-with-dips analysis, we fix the New Physics parameters to the best-fit values from the HESE-only analysis (see \Cref{tab:bf}). The results are presented in \Cref{fig:NT}, where we show the test-statistic for the two scenarios as well as our predicted event distributions. As can be seen, Northern Tracks are largely insensitive to the dips in the astrophysical neutrino spectrum. This insensitivity arises because the observed muons are predominantly through-going, and thus the observed muon energy does not correspond uniquely to the incident neutrino energy\footnote{A neutrino with larger energy interacting further from the detector and a neutrino will lesser energy interacting closer to the detector lead to a through-going muon of similar energy.}. As a result, any sharp spectral features are smeared out in the observed muon spectrum, and leave only a modest imprint on the event rates. The detector resolution (see \Cref{eq:Gauss}) leads to additional smearing for both hypotheses equally and is not the dominant cause of this insensitivity. As a result, the prediction for spectral index is indistinguishable for the two scenarios (see \Cref{fig:NT}, \emph{left}). This is also evident in our comparison of the event distributions for the the best-fit values for the two hypothesis (see \Cref{fig:NT}, \emph{right}). The event rates are statistically indistinguishable. The predicted normalizations for the two fits are slightly different. As mentioned earlier, our simplified treatment does not yield the correct normalisation and hence we do not report them. Due to the insensitivity of Northern Tracks data to the narrow spectral features, changing the location of the dips (i.e. a scan over $M_X$ and $m_s$) does not yield meaningful insights. On the other hand, broad spectral features (e.g., those considered in \cite{Fiorillo:2022rft} and \cite{Mohanty:2018cmq}) can lead to more pronounced differences and therefore are more testable with Northern Tracks.

\subsection{ESTES Analysis}

In \Cref{Sec:Methods-ESTES}, we outlined our methodology for estimating the distribution of ESTES events in IceCube. Similar to Northern Tracks analysis, we now compare our predictions with observed IceCube data to extract the astrophysical flux parameters as well as determine the sensitivity to dips in the spectrum. We again find that our simple methods systematically underestimate the ESTES event rates when compared to IceCube's detailed simulations. Similar to northern tracks analysis, we fix the atmospheric neutrino background and use only the high energy bins ($>70\,$TeV) in the likelihood. Again, the results in this section should be interpreted as trends and relative comparisons rather than as precise determinations.

\begin{figure}[h!]
    \centering
    \includegraphics[width=0.40\linewidth]{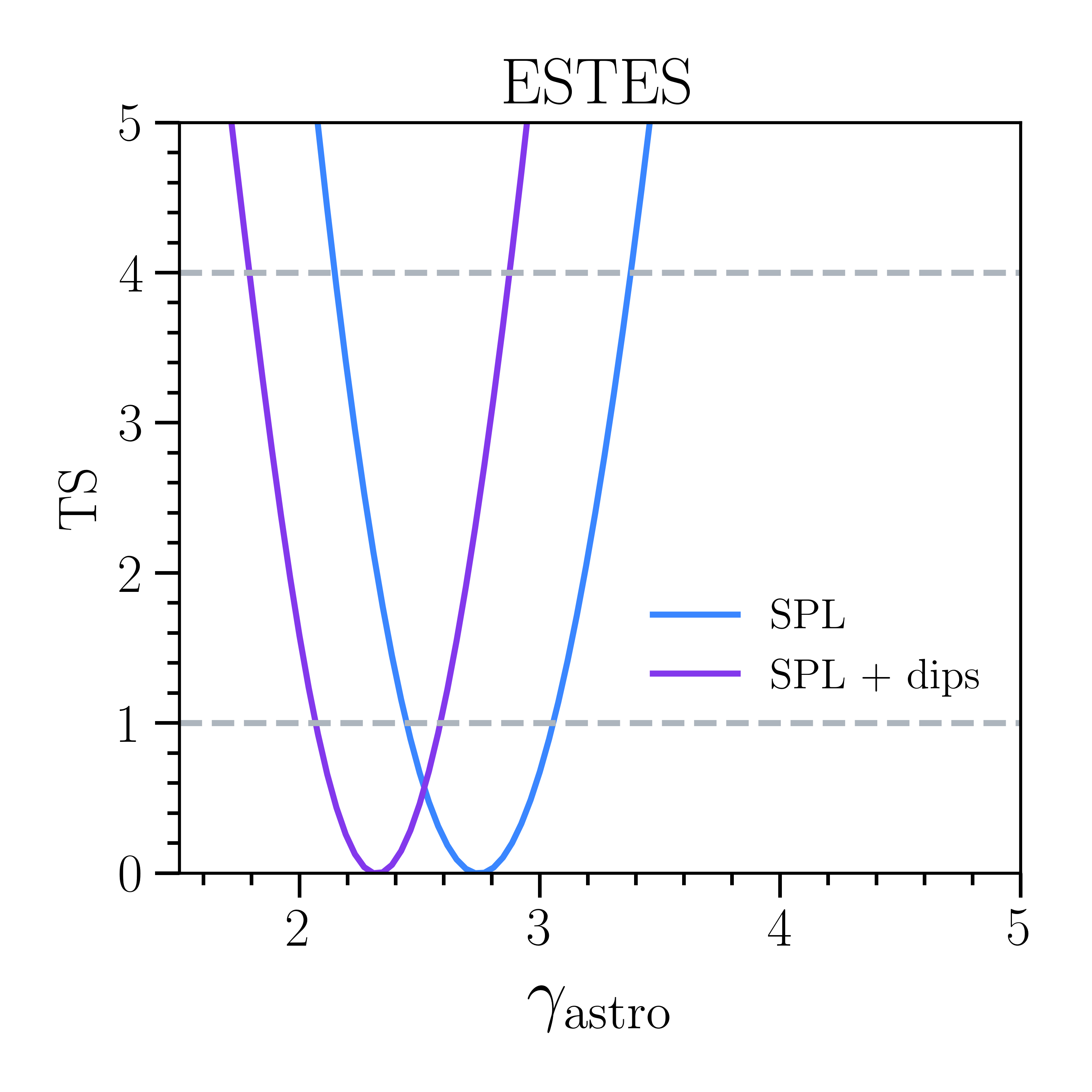}
    \includegraphics[width=0.56\linewidth]{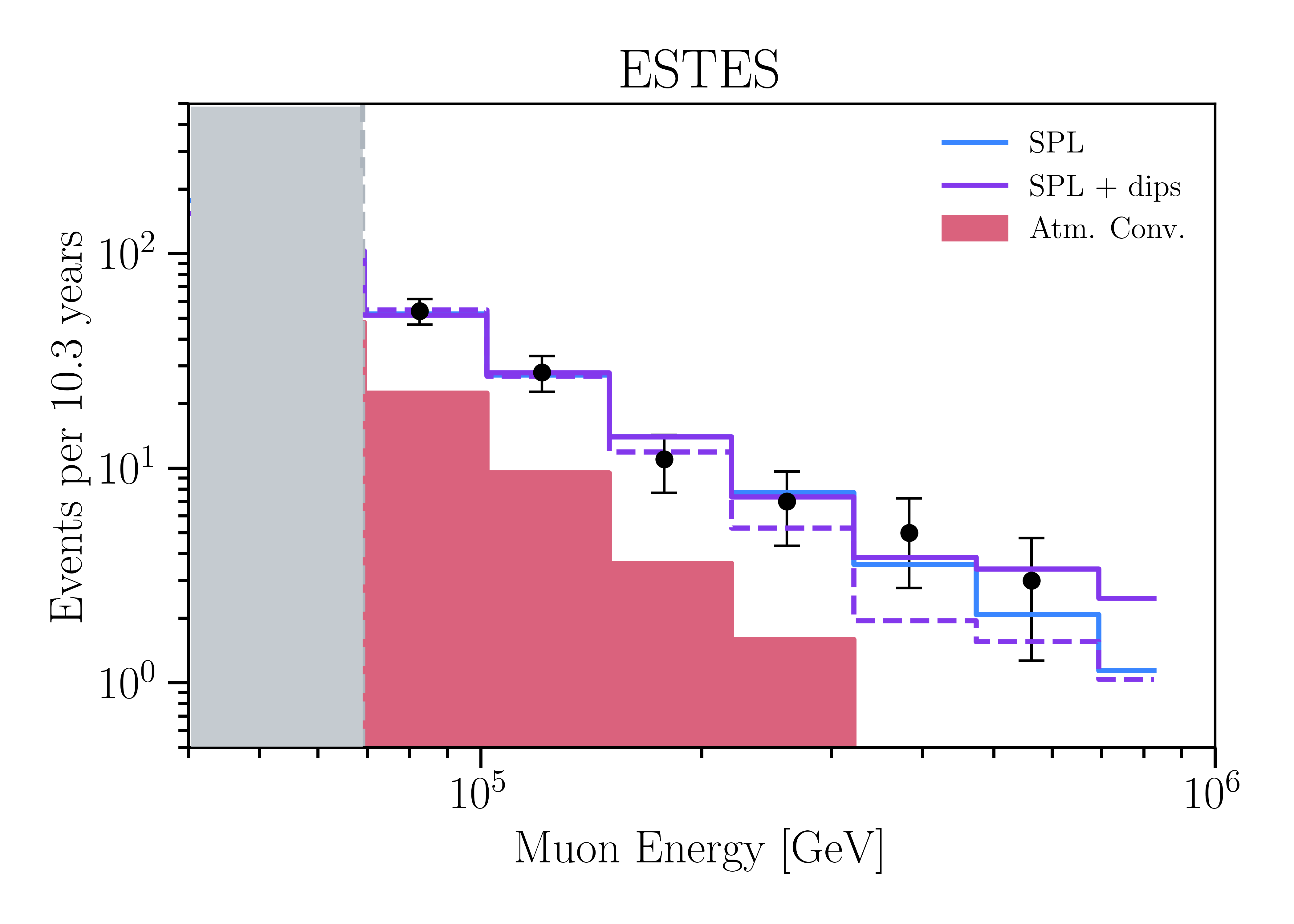}
    \caption{
    \emph{Left:} The test statistic (see \Cref{eq:ts}) for the single power-law fit of 10.5 years ESTES sample \cite{IceCube:2024fxo} is shown in blue. In purple, we show the test statistic after including absorption features using the best-fit New Physics parameters from \Cref{tab:bf}. The horizontal grey dashed lines are used to denote the 68\% C.L. and 90\% C.L. intervals. \emph{Right:} The event rates obtained using \Cref{eq:Gauss} for the SPL and SPL+dips are shown in blue and purple respectively. Note that the normalization for both fits is different. The purple dashed curve is obtained by considering dips with flux parameters from best-fit values of SPL, and shown only for comparison. The observed data points are taken from \cite{IceCube:2024fxo} and the errors are statistical only. The red shaded region is the atmospheric contribution. In the grey shaded region, the atmospheric contribution is large and hence excluded from our analysis (see text for details).
    }
    
    \label{fig:ESTES}
\end{figure}

Our ESTES analysis is similar to Northern Tracks and the results are presented in \Cref{fig:ESTES}. Since these are track-like events that start inside the detector, there is a better relationship between the measured muon energy and the incident neutrino energy. As a result, any narrow spectral features lead to difference in the predicted spectral index (see \Cref{fig:ESTES}, \emph{left}). The best fit values from our analysis are, 
\begin{equation}
    \hat{\gamma}^{\text{\,ESTES, SPL}}_{\texttt{\,astro}} = 2.74,\text{ and } \hat{\gamma}^{\text{\,ESTES, SPL+dips}}_{\texttt{\,astro}} = 2.32.
\end{equation}
The lowering of the spectral index is comparable to the results from HESE-only analysis.

We also compare the event distributions for the the best-fit values for the two hypothesis in \Cref{fig:NT} (\emph{right}). The predicted normalizations for the two fits are different. In addition, we also show the event rates obtained by considering dips and the SPL best-fit values for spectral index and normalization for comparison.  

\subsection{Forecast for IceCube-Gen2}

\begin{figure}[b]
    \centering
    \includegraphics[width=0.6\linewidth]{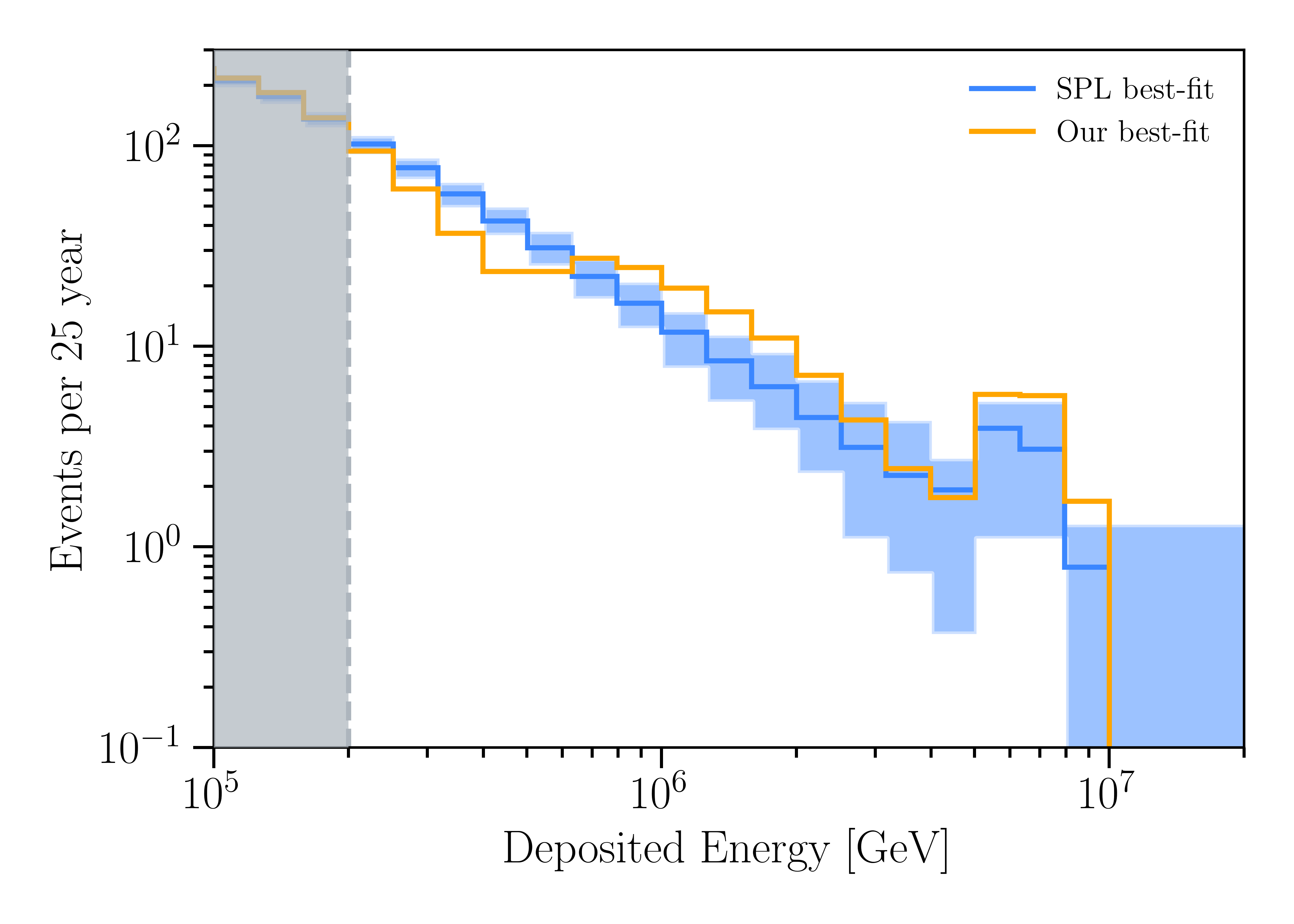}
    \caption{Event rate prediction for 25 years of exposure for IceCube-Gen2. The orange curve represents events assuming the best-fit parameters of this work (see \Cref{tab:bf}). The blue curve represents events for the SPL best-fit parameters. The blue-shaded region shows 68 \% C.L. errors on the SPL event rate assuming Poisson distribution. The grey-shaded region represents the 200 TeV threshold for IceCube-Gen2 \cite{IceCube-Gen2:2020qha}.}
    \label{fig:gen2}
\end{figure}

IceCube-Gen2 is the planned upgrade of the IceCube neutrino observatory, which is expected to have $\sim$10 times larger exposure to TeV--PeV neutrinos \cite{IceCube-Gen2:2020qha}. To estimate the event rate in IceCube-Gen2, we scale the weights according to the increased effective area. The effective area for cascades in IceCube-Gen2 is given in Ref. \cite{IceCube-Gen2:2020qha}. However, the effective area for tracks in IceCube-Gen2 is not publicly available at the time of writing this paper. In this work, we will assume that the effective area is the same for cascades and tracks. From the preceeding discussion, it is clear that starting events (HESE or ESTES) are good probes of dips in the spectrum. We have evaluated the HESE event rates using 25 years of IceCube-Gen2 exposure for the best-fit parameters from the SPL analysis and for the best-fit New Physics parameters from this work. The results are presented in \Cref{fig:gen2}. The local significance of the dips can be tested with such large exposures. It must be noted that there are other proposed neutrino telescopes that will also be sensitive to the absorption features. A combined analysis will improve the discovery prospects\,\cite{Fiorillo:2022rft}.

\section{Summary and outlook}\label{sec:summary}

The IceCube neutrino observatory has played a crucial role in our understanding of astrophysical neutrinos in the last decade. It also provides an opportunity to study rare interactions over large distances. In this paper, we have taken a close look at the interaction of the astrophysical neutrinos with a hypothetical sterile neutrino background from the perspective of various IceCube datasets. The light sterile neutrino is well-motivated from short baseline experiments. However, there is a conflict with standard cosmology, and one of the resolutions is to allow self-interactions via a MeV-scale mediator. As a consequence, one expects the cosmic neutrino background to have a sterile neutrino component as well. The astrophysical neutrinos can interact with the sterile neutrino background, which leads to dips in the TeV--PeV neutrino energy spectrum. If the dips are significant, there is an observable impact on the starting events such as HESE and ESTES. The Northern Tracks are not significantly affected as there is a lack of one-to-one correspondence between observed event energy and the incident neutrino energy.

We have improved on the previous HESE analysis on multiple fronts as well. We have taken a realistic source distribution and included the effect of regeneration during the propagation. We also account for the recent observation of Glashow resonance in our analysis. To estimate the likelihood, we first compute the event rate and then compare with the observations. We find that the inclusion of these absorption features allows for a lower spectral index ($\gamma_{\texttt{astro}} = 2.60^{+0.19}_{-0.16}$) when compared with the single-power law flux ($\gamma_{\texttt{astro}} = 2.87^{+0.20}_{-0.19}$). Note that, due to the approximations in our analysis, our errors are underestimated. The lower spectral index is in better agreement with the Northern Tracks sample ($\gamma_{\texttt{astro}} = 2.37^{+0.09}_{-0.09}$), which are largely insensitive to the presence of such narrow features in the neutrino spectrum. The ESTES sample can prove complementary probe of these dips and can provide information about the flavor structure, if any. Currently, with only 7.5 years of HESE data, we find that there is a large overlap of parameter space, and one cannot statistically infer the existence of such dips. In the future, with a much larger neutrino telescope in the form of IceCube-Gen2 and perhaps other IceCube-scale experiments, the absorption features will be strongly tested.

\section*{Acknowledgments}

The authors would like to thank Austin Schneider for helping with the HESE Monte Carlo simulation and Ivan Esteban for helping with nuSIprop. The authors would also like to thank Subhendra Mohanty and Ivan Esteban for their comments and suggestions on the manuscript. The authors thank Ranjan Laha for useful discussions. The authors acknowledge the anonymous referee who motivated a large body of work in this manuscript. BC is supported by an OPERA grant and seed grant NFSG/PIL/2023/P3821 from BITS Pilani. PP acknowledges the support from the National Post-Doctoral Fellowship by the Science and Engineering Research Board (SERB), Department of Science and Technology (DST), Government of India (SERB-PDF/2023/002356) and the IOE-IISc fellowship program. The authors acknowledge the use of computational resources provided by DIST-FIST Project No. SR/FST/PS-1/2017/30.

\bibliography{sisnu_references.bib}

\newpage
\section*{Appendix A}\label{sec:AppA}
\begin{figure}[h!]
    \centering
    \includegraphics[width=0.5\linewidth]{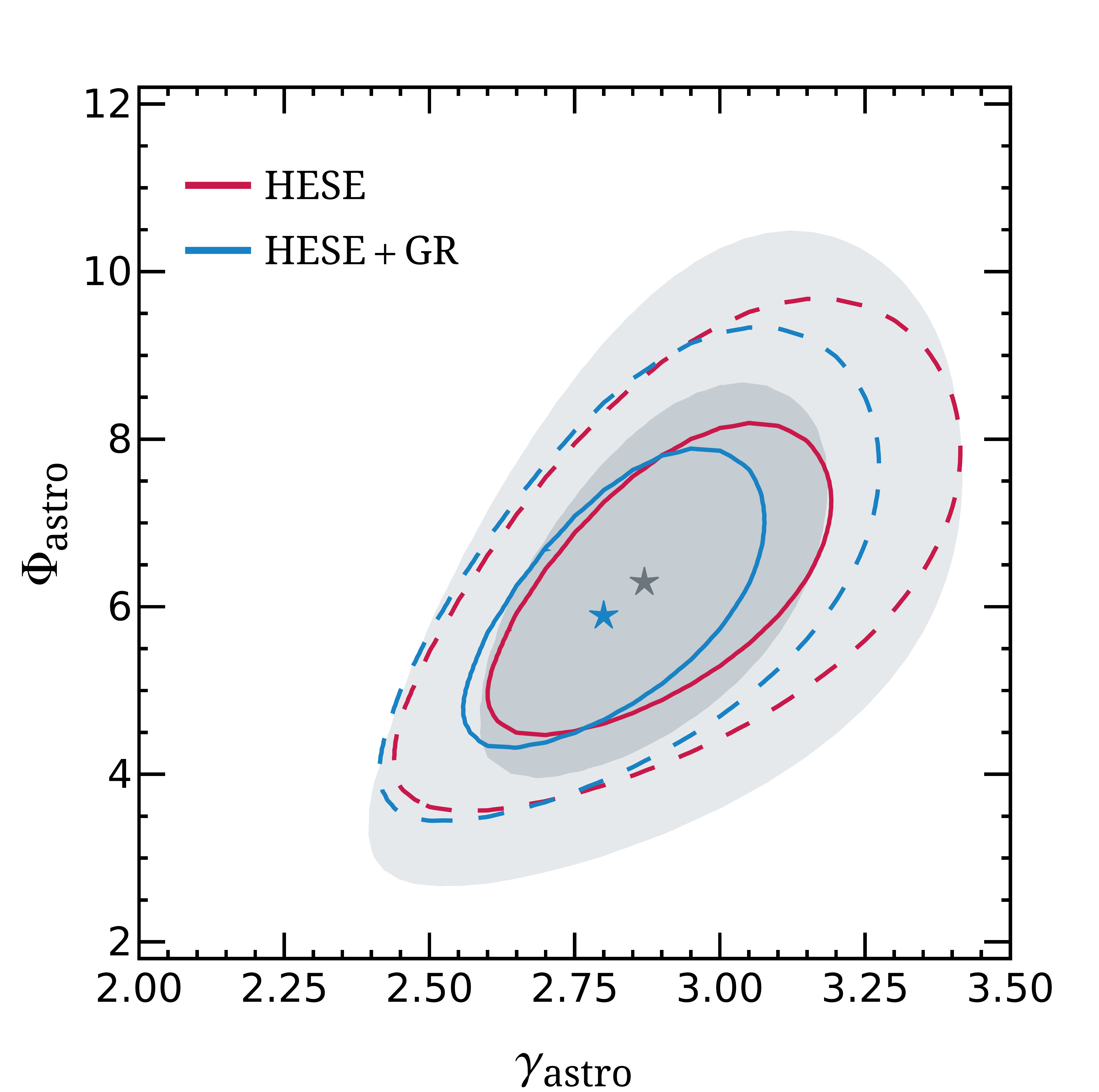}
    \caption{The SPL fit to HESE data \cite{IceCube:2020wum} considering all nuisance parameters (see \Cref{tab:params}) is shown in grey. The inner (dark grey) region shows the 68 \% C.L. interval and the outer (light grey) region shows the 90 \% C.L. interval. The best-fit point is shown as a grey star. Our fit to HESE data in which the nuisance parameters are fixed to their nominal values is shown as red curves. Our best-fit point for HESE-only analysis is very close to the IceCube best-fit point and not shown here. The fit including HESE and Glashow Resonance (GR) is shown in blue. The blue star shows the best-fit point of our HESE+GR analysis.}
    \label{fig:appA}
\end{figure}

In order to estimate the effects of our approximations on the fit, we evaluate the 68\% C.L. and 90\% C.L. intervals for single-power law and compare them with the contours in Ref.\,\cite{IceCube:2020wum}. The results are shown in \Cref{fig:appA}. We only scan over the astrophysical flux parameters and keep other parameters fixed close to their best-fit values. We find that on expected lines, the best-fit point agrees very well, and the errors are underestimated. It is also seen that the impact of these approximations lead to a larger change in the normalization and the impact on the spectral index is small. 

The impact of including the Glashow Resonance event is also on expected lines. In order to get a reasonable expectation for the energy bin, one must have a harder spectrum, and larger values of spectral index are not preferable. As one can see in the \Cref{fig:appA}, the fits for smaller values of $\gamma_{\texttt{astro}}$ for both HESE and HESE+GR are similar, and differ for larger values.

\end{document}